\documentclass[useAMS,usenatbib]{mn2e}
\usepackage{graphicx}
\usepackage{color}
\usepackage{amsmath}

\title[A treatment procedure for SINFONI data cubes]{A treatment procedure for VLT/SINFONI data cubes: application to NGC 5643}

\author[Menezes et al.] {R.~B.~Menezes$^1$\thanks{E-mail:
robertobm@astro.iag.usp.br}, Patr\'icia~da~Silva$^1$, T.~V.~Ricci$^1$, J.~E.~Steiner$^1$, D.~May$^1$, 
\newauthor and B.~W.~Borges$^2$ \\
$^{1}$Instituto de Astronomia Geof\'isica e Ci\^encias Atmosf\'ericas, Universidade de S\~ao Paulo, Rua do Mat\~ao 1226, \\Cidade Universit\'aria, S\~ao Paulo, SP CEP 05508-090, Brazil \\
$^{2}$Campus Ararangu\'a, Universidade Federal de Santa Catarina, Ararangu\'a, SC, CEP 88905-120, Brazil}

\begin{document}

\date{Accepted 2015 March 23. Received 2015 March 20; in original form 2014 August 25}

\pagerange{\pageref{firstpage}--\pageref{lastpage}} \pubyear{2015}

\maketitle

\label{firstpage}

\begin{abstract}

In this second paper of a series, we present a treatment procedure for data cubes obtained with the Spectrograph for Integral Field Observations in the Near Infrared of the \textit{Very Large Telescope}. We verified that the treatment procedure improves significantly the quality of the images of the data cubes, allowing a more detailed analysis. The images of the Br$\gamma$ and H$_2 \lambda 21218$ emission lines from the treated data cube of the nuclear region of NGC 5643 reveal the existence of ionized and molecular-gas clouds around the nucleus, which cannot be seen clearly in the images from the non-treated data cube of this galaxy. The ionized-gas clouds represent the narrow-line region, in the form of a bicone. We observe a good correspondence between the positions of the ionized-gas clouds in the Br$\gamma$ image and in an [O III] image, obtained with the \textit{Hubble Space Telescope}, of the nuclear region of this galaxy convolved with an estimate of the point-spread function of the data cube of NGC 5643. The morphologies of the ionized and molecular gas seem to be compatible with the existence of a molecular torus/disc that collimates the active galactic nucleus (AGN) emission. The molecular gas may also flow along this torus/disc, feeding the AGN. This scenario is compatible with the unified model for AGNs.

\end{abstract}


\section{Introduction}

\begin{table*}
\begin{center}
\caption{Data used throughout the paper to present the treatment procedure.\label{tbl1}}
\begin{tabular}{cccccccc}
\hline
Target  &  Programme ID  &  Name of PI  &  Observation date  &  Spaxel scale (mas)  &  Filter  &  Exposure time (s)  &  Number \\
            &                           &                      &                               &                                  &            &                                &  of frames \\
\hline
HIP 044245  &  60.A-9235(A)    &          -          &        2009 April 8        &  250  &  \textit{K}       &    2    &  1  \\
HIP 048480  &  074.B-9012(A)  &  Eisenhauer  &      2005 March 16     &  100  &  \textit{K}       &   30   &  1  \\
HIP 049220  &  074.B-9012(A)  &  Eisenhauer  &  2004 December 22   &   25   &  \textit{K}       &   10   &  1  \\
HIP 096155  &   60.A-9235(A)   &          -          &    2007 August 14      &  250  &  \textit{H+K}  &  1.5   &  1  \\
HIP 091481  &   60.A-9235(A)   &          -          &  2006 September 17  &  100  &  \textit{H+K}  &   30   &  1  \\
HIP 024761  &   60.A-9235(A)   &          -          &      2011 August 6      &   25   &  \textit{H+K}  &    2    &  1  \\
NGC 5128    &  074.A-9011(A)   &  Eisenhauer  &      2005 March 24     &  100  &  \textit{K}       &  900  &  9  \\
NGC 5128    &  075.B-0490(A)   &   Cappellari  &          2005 April 2       &  250  &  \textit{J}       &  300  &  4  \\
NGC 5128    &  075.B-0490(A)   &   Cappellari  &          2005 April 2       &  250  &  \textit{H}      &  300  &  4  \\
NGC 5128    &  075.B-0490(A)   &   Cappellari  &          2005 April 2       &  250  &  \textit{K}      &  300  &  4  \\
NGC 5643    &  083.B-0332(A)   &       Hicks      &          2009 April 8       &  250  &  \textit{K}      &  600  &  5  \\
\hline
\end{tabular}
\end{center}
\end{table*}

Data cubes are data sets with two spatial dimensions and one spectral dimension. Data cubes enable the analysis of spectra at different spatial positions of a given object, or to perform cuts along the spectral direction, and thus to produce images of the object at different wavelengths. The analysis of data cubes in astronomy in the last decades has increased significantly with the construction of more instruments like Integral Field Spectrographs and Fabry-Perot spectrographs. Although the wealth of information in this kind of data is very significant, the performed analysis is usually affected by the existence of artefacts, like high spatial-frequency noise or `instrumental fingerprints' (see Fig. 5 of Neumayer et al. 2007). Therefore, a proper removal of these artefacts is recommended if one wishes to perform a cleaner analysis. 

The Spectrograph for Integral Field Observations in the Near Infrared (SINFONI; Eisenhauer et al. 2003; Bonnet et al. 2004) of the \textit{Very Large Telescope} (\textit{VLT}) provides 3D imaging spectroscopy in a spectral range of $1.05 - 2.45 \mu$m (covering the spectral bands \textit{J}, \textit{H} and \textit{K}). This instrument, actually, consists of a combination of the Adaptive Optics Module, developed by the \textit{European Southern Observatory}, with the Spectrometer for Infrared Faint Field Imaging (SPIFFI), developed by the Max-Planck-Institute for extraterrestrial Physics. SINFONI has three different fore-optics, which result in field of views (FOVs) of 8.0 arcsec $\times$ 8.0 arcsec (hereafter 8.0 arcsec), 3.2 arcsec $\times$ 3.2 arcsec (hereafter 3.2 arcsec) and 0.8 arcsec $\times$ 0.8 arcsec (hereafter 0.8 arcsec). These three fore-optics have spatial pixels (spaxels) of 0.125 arcsec $\times$ 0.250 arcsec (spaxel scale of 250 mas), 0.05 arcsec $\times$ 0.10 arcsec (spaxel scale of 100 mas) and 0.0125 arcsec $\times$ 0.0250 arcsec (spaxel scale of 25 mas), respectively. The fore-optics with an FOV of 0.8 arcsec allows observations with a spatial resolution of about 0.056 arcsec in a wavelength of $2.2 \mu$m (which is a near-diffraction limited resolution), when the adaptive optics (AO) is applied. The SINFONI detector is composed by one CCD with 2048 $\times$ 2048 pixels.

The incident light in the FOV of SINFONI passes through the pre-optics region, which has three main sub-areas. In the first of these sub-areas, the background thermal emission is suppressed. The second sub-area is a motorized filter wheel that selects one of the four broad-band filters available: \textit{J}, \textit{H}, \textit{K} and \textit{H+K}. Since the \textit{H+K} filter corresponds to the \textit{H} and \textit{K} bands together, it provides the largest spectral coverage, but the lowest spectral resolution. The third sub-area is the motorized optics wheel, which selects one of three lenses corresponding to the available fore-optics (with FOVs of 8.0, 3.2 and 0.8 arcsec). After passing through the pre-optics region, the light is transferred to the image slicer, where, at first, it is divided into several parts (the slitlets) by a group of 32 small mirrors (the small slicer). After this division, the light  passes by a second group of big mirrors (the big slicer), which reconfigure all the slitlets in a way that, at the end of this process, they are all disposed in a shape similar to a slit. The light is then transferred to the collimator, where it is redirected to the region containing the diffraction gratings (which generate the spectra). Finally, the light is transferred to the detector of the instrument. Each of the 32 slices is projected on to 64 detector pixels, explaining why the spaxels on the sky are rectangular.

\begin{figure*}
\begin{center}
  \includegraphics[scale=0.65]{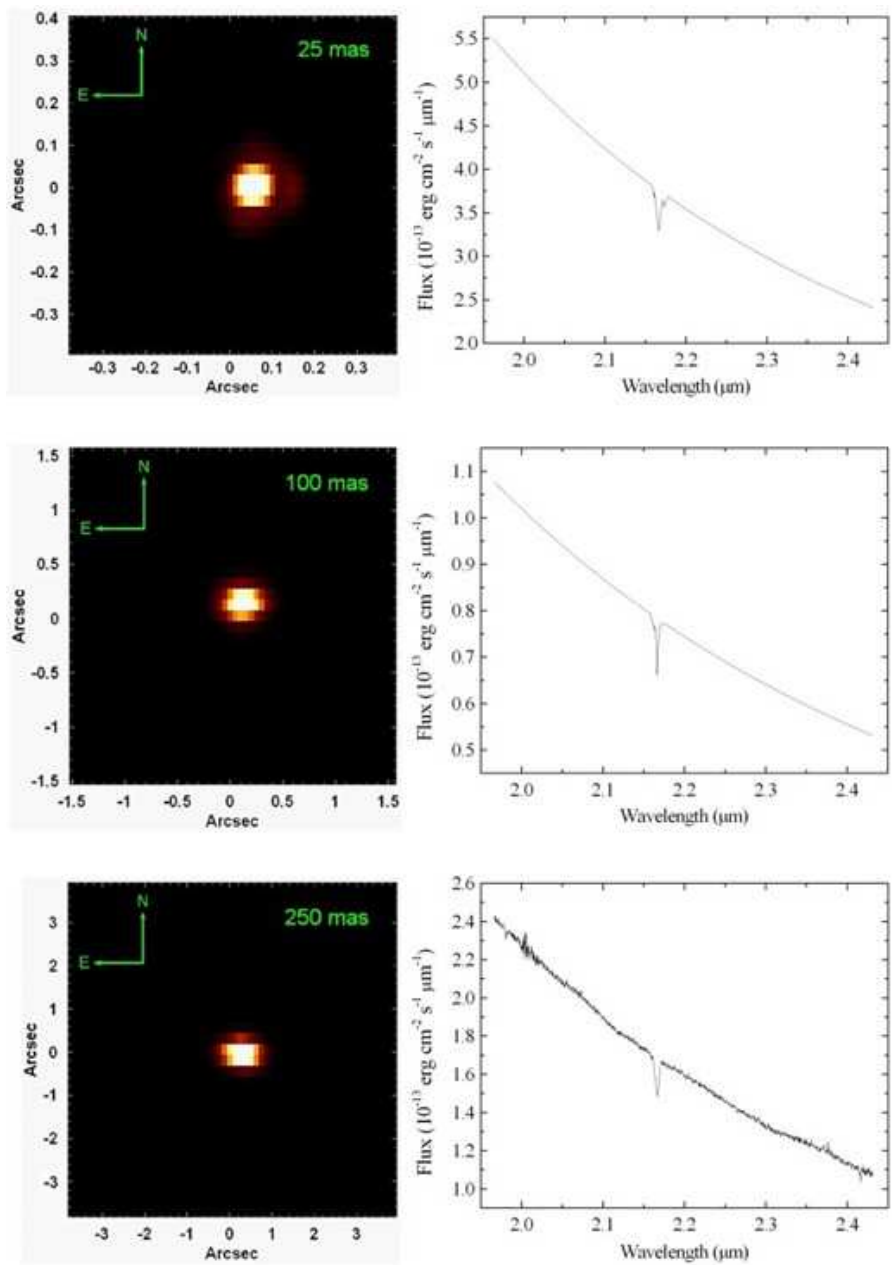}
  \caption{Average spectra and images of intermediate-wavelength intervals of the data cubes of (top) HIP 049220, (middle) HIP 048480 and (bottom) HIP 044245, obtained after data reduction, using the standard pipeline. Note that the contours of the stars are not well defined in the images. In addition, high spatial-frequency components can also be observed.\label{fig1}}
\end{center}
\end{figure*}

This is our second paper focused on describing data treatment procedures for data cubes obtained with different instruments. In Menezes et al. (2014, hereafter Paper I), we presented a procedure for treating data cubes (after the data reduction) obtained with the Near-Infrared Integral Field Spectrograph (NIFS; McGregor et al. 2002). In this paper, we present a similar procedure for treating SINFONI data cubes. Although the procedures described here and in Paper I have essentially the same steps, there are many differences in the way each one of these steps should be applied to NIFS and to SINFONI data cubes, in order to obtain optimized results. All the steps in this sequence were applied using scripts written in Interactive Data Language\footnote{http://www.exelisvis.com/ProductsServices/IDL.aspx} ({\sc idl}). The data used in the paper to illustrate the treatment procedure are described in Section 2. In Section 3, we present details about the correction of the differential atmospheric refraction (DAR). In Section 4, we describe the spatial re-sampling of the data. The Butterworth spatial filtering, to remove high frequency components from the images of the data cubes, is discussed in Section 5. The process of `instrumental fingerprint' removal is presented in Section 6. The Richardson-Lucy deconvolution, to improve the spatial resolution of the data cubes, is described in Section 7. The analysis of the data cubes of NGC 5643 (used as a scientific example to show the benefits provided by the data treatment) is discussed in Section 8. Finally, in Section 9, we present a summary of the procedure and our conclusions.

\section{Observations and data reduction}

To illustrate most of the treatment procedure, we used observations of the standard stars HIP 044245, HIP 048480 and HIP 049220, available in the SINFONI public access data archive\footnote{http://archive.eso.org/wdb/wdb/eso/sinfoni/form}. All the observations were taken in the \textit{K} band, with different fore-optics. 

For the analysis of the spatial displacements of the objects along the spectral axis of SINFONI data cubes (Section 3), we used observations of the standard stars HIP 096155, HIP 091481 and HIP 024761, which were taken in the \textit{H+K} band, with different fore-optics. We chose observations of stars in this specific spectral band because it has the highest spectral coverage, which makes it easier to analyse the spatial displacements along the spectral axis of the data cubes. To show the process for `instrumental fingerprint' removal (Section 6), we used observations of the nuclear region of the galaxy NGC 5128 (Centaurus A), in the \textit{J}, \textit{H} and \textit{K} bands in different fore-optics. 

Finally, as a scientific example illustrating the benefits obtained with the data treatment (Section 8), we analysed observations of the nuclear region of the galaxy NGC 5643, in the \textit{K} band. The AO was applied to all the observations taken with the fore-optics with FOVs of 3.2 and 0.8 arcsec; however, it was not applied to the observations taken with the fore-optics with FOV of 8.0 arcsec. The details of all the data used throughout this paper are shown in Table~\ref{tbl1}. 

\begin{figure*}
\begin{center}
  \includegraphics[scale=0.43]{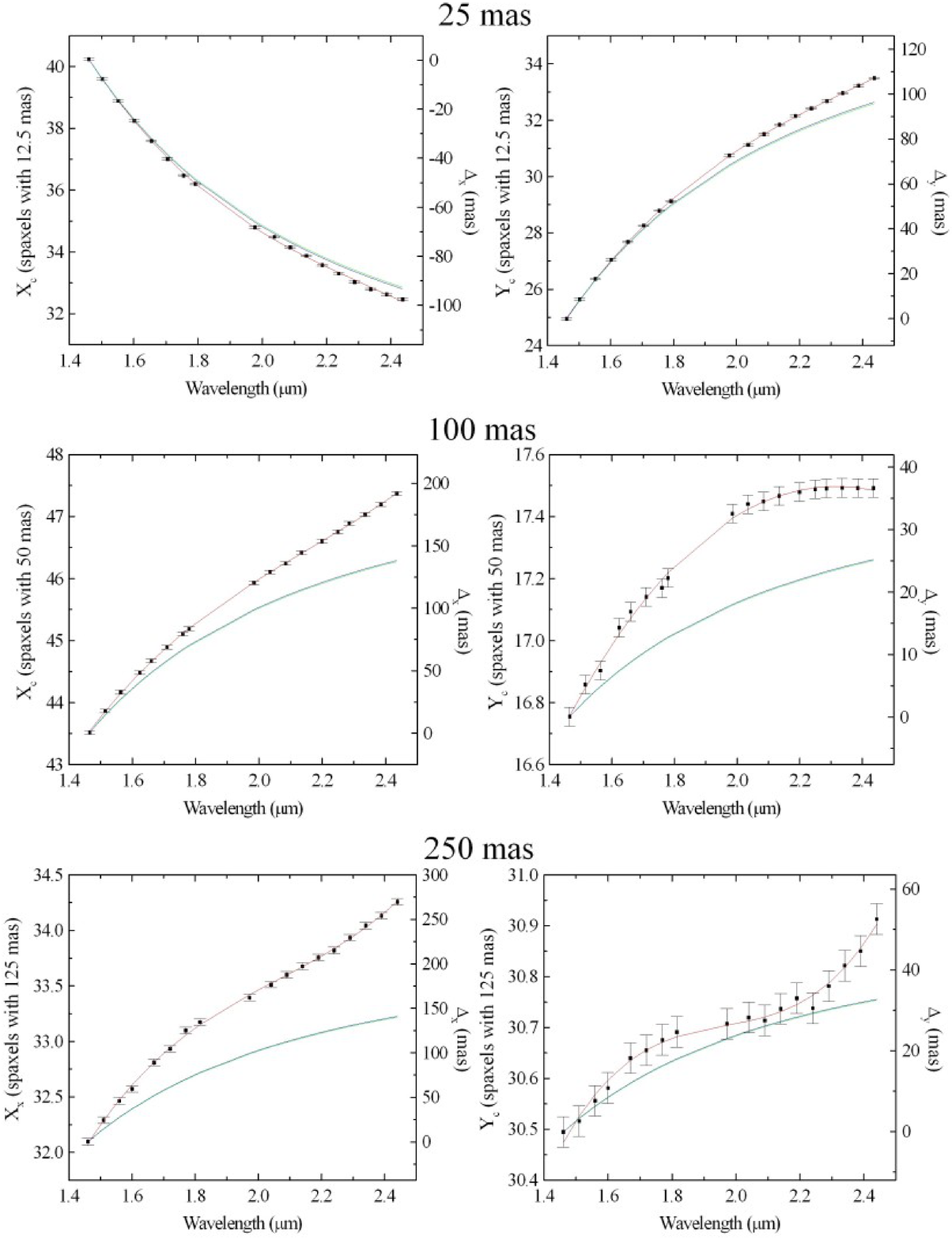}
  \caption{Graphs of the coordinates $X_c$ and $Y_c$ of the centres of the stars (top) HIP 024761, (middle) HIP 091481 and (bottom) HIP 096155 as a function of the wavelength. The values of $\Delta_x$ and $\Delta_y$ represent the spatial shifts along the horizontal and vertical axis, respectively, from the positions of the stars at the blue extremities of the data cubes. The red curves are third degree polynomials, the green curves were obtained with the equations from \citet{bon98} and the blue curves were obtained with the task `\textit{refro}' from the {\sc`slalib'} package. The green and blue curves are almost coincident; therefore, it is very difficult to differentiate them.\label{fig2}}
\end{center}
\end{figure*}

\begin{figure}
\begin{center}
  \includegraphics[scale=0.44]{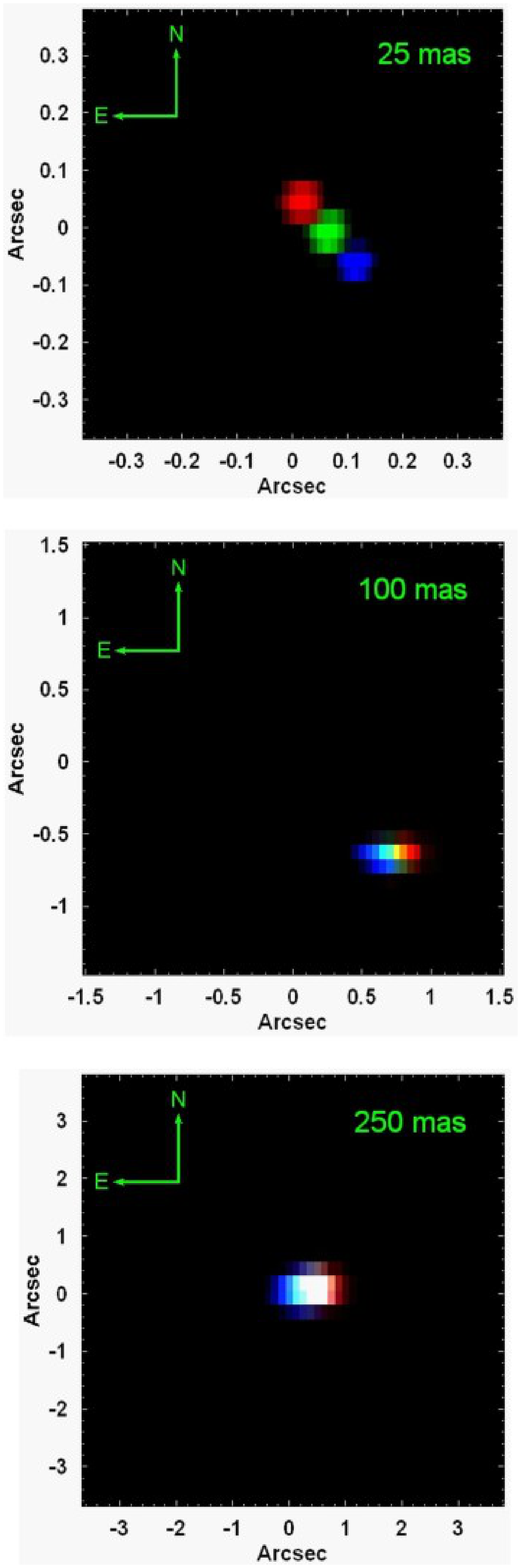}
  \caption{RGB composite images of the data cubes of the stars (top) HIP 024761, (middle) HIP 091481 and (bottom) HIP 096155. In all images, the blue colour corresponds to the wavelength range 1.4615 - 1.4685 $\mu$m, the green colour corresponds to the wavelength range 1.7920 - 1.7990 $\mu$m and the red colour corresponds to the wavelength range 2.4485 - 2.4555 $\mu$m.\label{fig3}}
\end{center}
\end{figure}

Standard calibration data of linearity lamp, distortion fibre (used in the data reduction to compute spatial distortions and to perform a spatial rectification), flat lamp, arc lamp and sky field were obtained from the public data archive. Calibration spectra of standard stars were also retrieved for the reduction of the data cubes of NGC 5128 and NGC 5643. The data reduction was performed with the {\sc gasgano} software\footnote{http://www.eso.org/sci/software/gasgano.html} and included the following steps: correction of bad pixels, flat-field correction, spatial rectification (including correction for spatial distortions), wavelength calibration, sky subtraction and data cube construction. One data cube was obtained for each one of the following standard stars: HIP 044245, HIP 048480, HIP 049220, HIP 096155, HIP 091481 and HIP 024761. The data cubes of HIP 044245 and HIP 096155 were reduced with spaxels of 0.125 arcsec $\times$ 0.125 arcsec (hereafter 125 mas), the data cubes of HIP 048480 and HIP 091481 were reduced with spaxels of 0.05 arcsec $\times$ 0.05 arcsec (hereafter 50 mas) and the data cubes of HIP 049220 and HIP 024761 were reduced with spaxels of 0.0125 arcsec $\times$ 0.0125 arcsec (hereafter 12.5 mas). Throughout this paper, we will refer mainly to the original spaxel scales (25, 100 and 250 mas) of the data cubes of these standard stars (not to their HIP identifiers), as this is the most important factor in our discussion. Nine data cubes were obtained for NGC 5128, with spaxels of 50 mas, and five data cubes for NGC 5643, with spaxels of 125 mas. Finally, all data cubes were flux calibrated and telluric corrected within {\sc iraf}\footnote{http://iraf.noao.edu}, using appropriate standard stars (assuming they have blackbody spectra). Fig.~\ref{fig1} shows images of intermediate-wavelength intervals of the data cubes of HIP 049220, HIP 048480 and HIP 044245, obtained after data reduction, using the standard pipeline, as well as the corresponding average spectra.

\section{Correction of the DAR}

The first step in our data treatment sequence is the correction of the DAR effect, which causes a displacement of the observed object along the spectral axis of a data cube. A detailed description of the DAR is given in Paper I. More details can also be found in \citet{bon98} and \citet{fil82}.

Before this procedure is discussed here, however, it is necessary to evaluate the significance of the DAR effect in SINFONI data cubes. This is particularly important in this case because, since the DAR is stronger in the optical than in the infrared, this effect is usually ignored in data cubes obtained in the infrared. In order to evaluate how significant is the spatial displacement of structures along the spectral axis of SINFONI data cubes, we determined the coordinates $X_c$ and $Y_c$ of the centres of the stars HIP 024761, HIP 091481 and HIP 096155, which were observed in the \textit{H+K} band, with spaxel scales of 25, 100 and 250 mas, respectively. Fig~\ref{fig2} shows the graphs of the coordinates $X_c$ and $Y_c$ of the centres of these stars as a function of wavelength. 

An analysis of Fig.~\ref{fig2} reveals that the total spatial displacements, along the \textit{H+K} band, of the data cubes with spaxel scales of 25, 100 and 250 mas are $\sim$100, $\sim$190 and $\sim$270 mas, respectively, along the horizontal axis, and $\sim$110, $\sim$40 and $\sim$50 mas, respectively, along the vertical axis. The total displacements of the star observed with spaxel scale of 25 mas along the horizontal and vertical axis are almost twice the predicted spatial resolution (56 mas, with the use of the AO) for the corresponding fore-optics (with an FOV of 0.8 arcsec). This indicates that these displacements are relevant for studies involving the \textit{H} and the \textit{K} bands together. However, the spatial displacements are comparable to the predicted spatial resolution even if only one of these spectral bands is used; therefore, we conclude that the DAR effect may be significant for SINFONI data cubes obtained with the fore-optics with an FOV of 0.8 arcsec. In the case of the star observed with spaxel scale of 100 mas, the spatial displacement along the horizontal axis was also almost twice the typical spatial resolution (100 mas, with the use of the AO) obtained with the corresponding fore-optics (with an FOV of 3.2 arcsec); therefore, such displacement is relevant for studies involving the \textit{H} and the \textit{K} bands together. However, similarly to what was observed with the star with spaxel scale of 25 mas, even the spatial displacement along one of these two spectral bands is still comparable to the typical spatial resolution of this fore-optics. Based on these observations, we conclude that the DAR effect may be also significant for SINFONI data cubes obtained with the fore-optics with an FOV of 3.2 arcsec. Finally, in the case of the star observed with spaxel scale of 250 mas, it is important to remember that the AO was not applied during the observation (this is usually the case when the corresponding fore-optics is used). Therefore, the predicted spatial resolution for this fore-optics depends fundamentally on the seeing conditions of the observations. Considering that, we can see that the observed spatial displacements of this data cube along the horizontal and vertical axis are smaller than the typical spatial resolutions obtained with this fore-optics. As a consequence, we conclude that the correction of the DAR effect may not be necessary in SINFONI data cubes obtained with the fore-optics with an FOV of 8.0 arcsec. 

The entire discussion above indicates that the correction of the DAR may be relevant in some cases, particularly when the fore-optics with FOVs of 0.8 and 3.2 arcsec are used. In Fig.~\ref{fig3}, we can see RGB 
composite images showing different wavelength intervals of the data cubes with spaxel scales of 25, 100 and 250 mas. The scientific necessity of the correction of the DAR in the data cubes with scales of 25 and 100 mas is evident in these RGBs. On the other hand, we can also see that the correction of the DAR may not be really necessary in the data cube with scale of 250 mas.

Fig.~\ref{fig2} revealed that the DAR may be important even when only one spectral band is analysed. However, the spatial displacements caused by this effect are obviously more significant when more than one band is taken into account. A study that often involves the entire continuum of the \textit{J}, \textit{H} and \textit{K} bands together is the spectral synthesis. Analysis of spectral 
lines may also be compromised by the DAR if the purpose is to compare images of lines located in different spectral regions. This is the case in the analysis of emission line ratios, for example.

\begin{figure}
\begin{center}
  \includegraphics[scale=0.45]{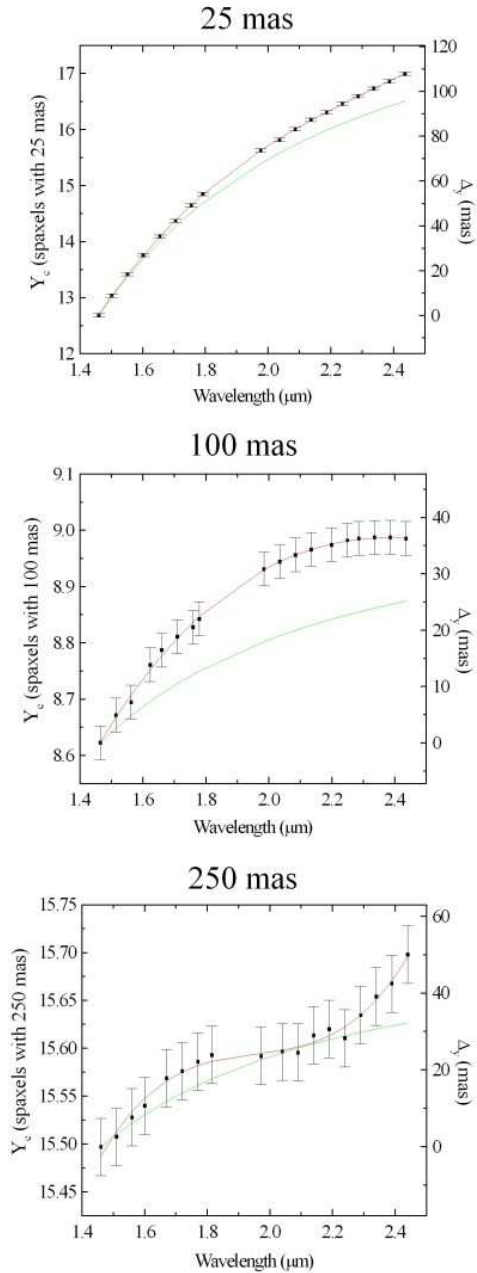}
  \caption{Graphs of the coordinates $Y_c$ of the centres of the stars (top) HIP 024761, (middle) HIP 091481 and (bottom) HIP 096155, keeping the original size of the spaxels along the vertical axis (25 mas for HIP 024761, 100 mas for HIP 091481 and 250 mas for HIP 096155). The values of $\Delta_y$ represent the spatial shifts along the vertical axis from the positions of the stars at the blue extremities of the data cubes. The red curves are third degree polynomials and the green curves were obtained with the equations from \citet{bon98}.\label{fig4}}
\end{center}
\end{figure}

The data cubes with spaxel scales of 25, 100 and 250 mas were observed with considerably high zenith distances: $62\degr\!\!.08$, $63\degr\!\!.28$ and $63\degr\!\!.56$, respectively, corresponding to airmasses of 2.14, 2.22 and 2.25, respectively. The DAR will be less significant for observations with lower zenith distances and, in such cases, even studies using observations taken with the fore-optics with FOVs of 0.8 and 3.2 arcsec or studies involving the continua in different spectral bands simultaneously may not require any correction.

\begin{figure*}
\begin{center}
  \includegraphics[scale=0.45]{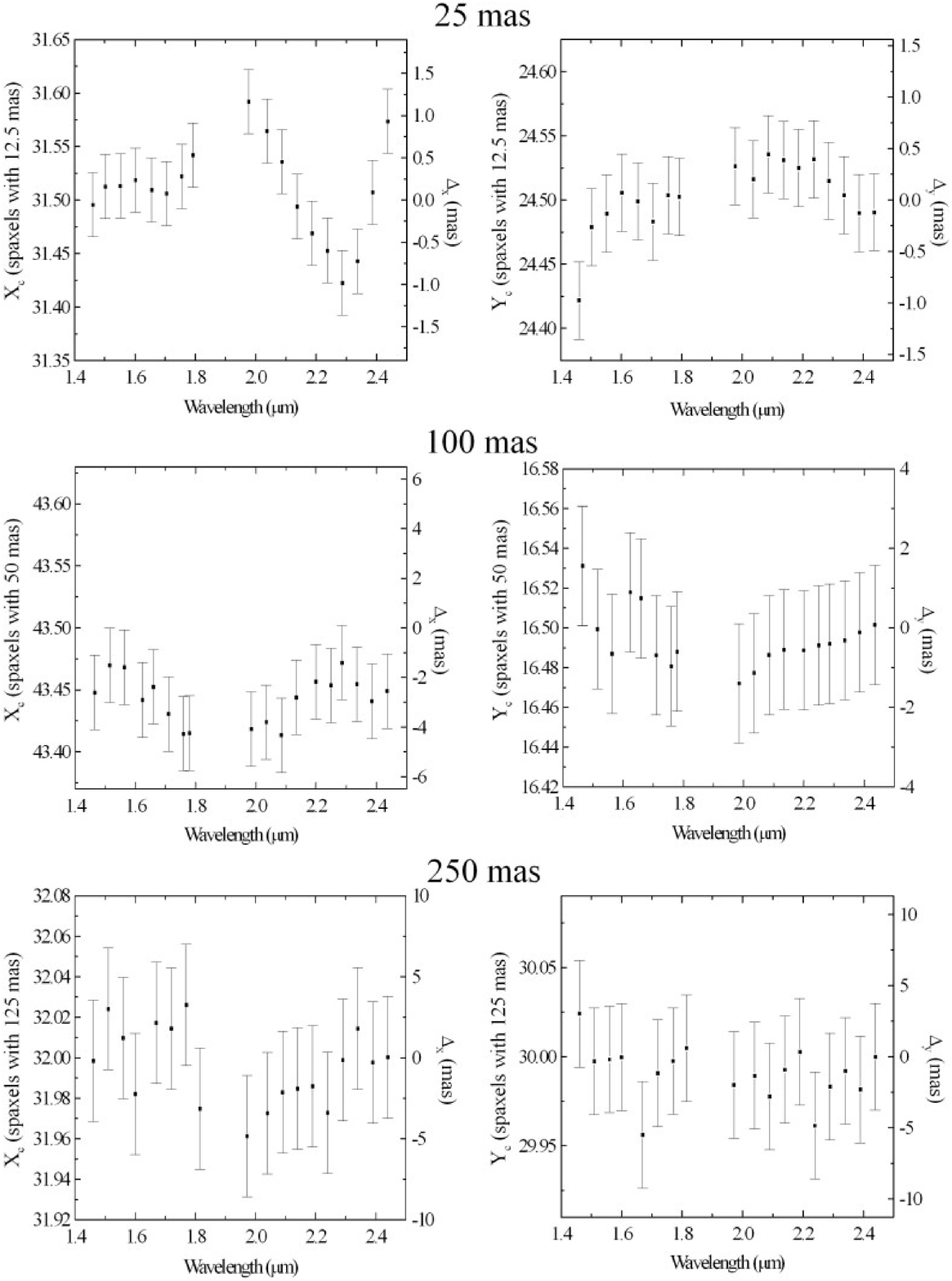}
  \caption{Graphs of the coordinates $X_c$ and $Y_c$ of the centres of the stars (top) HIP 024761, (middle) HIP 091481 and (bottom) HIP 096155 as a function of the wavelength, after the correction of the DAR, using the practical approach. The values of $\Delta_x$ and $\Delta_y$ represent the spatial shifts along the horizontal and vertical axis, respectively, from the expected positions of the stars, after the correction of the DAR.\label{fig5}}
\end{center}
\end{figure*}

Similarly to what we discussed in Paper I for NIFS data cubes, the significant impact of the DAR in some SINFONI data cubes may be somewhat unexpected, as this effect is usually ignored in infrared 3D spectroscopy. However, one should keep in mind that AO provides very high spatial resolutions to SINFONI data cubes. Therefore, although the DAR is indeed smaller in the infrared than in the optical, this effect may be relevant in SINFONI data cubes due to the high spatial resolutions. The DAR will have more impact on future data cubes obtained with the next generation of extremely large telescopes (of 25-39 m apertures), as the AO-corrected point-spread functions (PSFs) in these observations will be considerably smaller.

The script we use for the DAR correction shifts each image of the data cube in order to remove any spatial displacement of the structures along the spectral axis. This correction can be achieved using two strategies: the theoretical approach, in which the spatial displacements are calculated using the theoretical equations of \citet{bon98} and \citet{fil82}, and the practical approach, in which the spatial displacements are measured directly. These two strategies are described in further detail in Paper I.

The theoretical equations from \citet{bon98} assume a plane-parallel atmosphere. However, a more precise calculation is performed by the task `\textit{refro}' from the {\sc `slalib'} package, which takes into account the curvature of the Earth's atmosphere and an approximation to the change in atmospheric temperature and pressure with altitude. Fig.~\ref{fig2} shows third degree polynomials fitted to the points, the theoretical curves obtained with the equations from \citet{bon98} and the theoretical curves obtained with the task `\textit{refro}' from the {\sc`slalib'} package (implemented in Fortran environment).

Fig.~\ref{fig2} reveals no significant difference between the theoretical curves obtained with the equations from \citet{bon98} and with the task `\textit{refro}' from the {\sc `slalib'} package. These theoretical curves also do not reproduce exactly the values of $X_c$ and $Y_c$ of the centres of the stars along the spectral axis. The highest discrepancies between the observed values and the theoretical curves for the data cubes with spaxel scales of 25, 100 and 250 mas are $\sim$5, $\sim$55 and $\sim$130 mas, respectively, along the horizontal axis, and $\sim$11, $\sim$16 and $\sim$20 mas, respectively, along the vertical axis. These discrepancies are very common in SINFONI data cubes. Similar results were also obtained in Paper I for NIFS data cubes. Based on that, we conclude that the theoretical curves (considering a plane-parallel atmosphere or even taking into account the curvature of the atmosphere) are not the most precise way to reproduce the spatial displacements of structures along the spectral axis of SINFONI data cubes. As a consequence, the practical approach has a higher precision than the theoretical approach for removing the DAR from SINFONI data cubes. Nevertheless, one should keep in mind that the observed discrepancies mentioned before are all smaller than the predicted spatial resolutions of the observations in the corresponding fore-optics. We have never seen discrepancies much higher than these. Therefore, we can say that the theoretical approach is also a sufficiently precise way for removing the DAR from SINFONI data cubes. In some cases, there may be real physical effects that could cause the spatial displacement of an object along the spectral axis of a data cube. In such cases, the practical approach should not be used. 

It is important to try to determine the cause of the discrepancies between the observed values of $X_c$ and $Y_c$ and the theoretical curves of the DAR effect, detected in Fig.~\ref{fig2}. They were not caused by an imprecise calculation of the DAR, as even the detailed calculation performed by the task `\textit{refro}' did not reproduce the observed spatial displacements. As explained in Paper I, a possible cause for the discrepancies is the spatial re-sampling performed during the data reduction. Indeed, the data cubes obtained with the fore-optics with FOVs of 0.8, 3.2 and 8.0 arcsec were reduced with spaxels of 0.0125 arcsec $\times$ 0.0125 arcsec (hereafter 12.5 mas), 0.05 arcsec $\times$ 0.05 arcsec (hereafter 50 mas) and 0.125 arcsec $\times$ 0.125 arcsec (hereafter 125 mas), respectively; however, the original size of their spaxels were 0.0125 arcsec $\times$ 0.025 arcsec, 0.05 arcsec $\times$ 0.1 arcsec and 0.125 arcsec $\times$ 0.25 arcsec, respectively. Therefore, the data reduction did not alter the size of the spaxels along the horizontal axis but reduced by half their sizes along the vertical axis. To check if this spatial re-sampling affected significantly the values of $Y_c$ of the centres of the standard stars analysed here, we performed a new reduction of these data cubes, but keeping the size of the spaxels along the vertical axis unchanged. Then, we determined again the values of $Y_c$ of the centres of the stars along the spectral axis of the data cubes. Fig.~\ref{fig4} shows the results and it is easy to see that the previously observed discrepancies between the observed values of $Y_c$ and the theoretical curves remained essentially the same.

Based on Figs. 2 and 4, we can say that the discrepancies between the observed values of $X_c$ and $Y_c$ and the theoretical curves are not due to an imprecise calculation of the DAR effect or to the spatial re-sampling applied during the data reduction. We were unable to determine the exact cause of these discrepancies. We believe that an instrumental cause for this behaviour is a possibility. The fact that the observed discrepancies appear to increase from the fore-optics with an FOV of 0.8 arcsec to the fore-optics with an FOV of 8.0 arcsec for stars observed with very similar zenith distances seems to support the idea of an instrumental effect. A proof of this hypothesis, however, is beyond the scope of this work, due to the necessity of the analysis of many data cubes, obtained with different observational parameters. 

We applied the correction of the DAR, using the practical approach, to the data cubes of the standard stars with spaxel scales of 25 mas, 100 mas and 250 mas. The graphs with the values of $X_c$ and $Y_c$ of the centres of the stars in the obtained data cubes, as a function of the wavelength, are shown in Fig.~\ref{fig5}. It is easy to see that the procedure was very effective in removing the DAR. Since the expected values, after the correction, for $X_c$ and $Y_c$, respectively, were 31.5 and 24.5 for the data cube with spaxel scale of 25 mas, 43.5 and 16.5 for the data cube with spaxel scale of 100 mas and 32.0 and 30.0 for the data cube with spaxel scale of 250 mas, we can say that the precision ($1\sigma$) of our algorithm was higher than 3 mas in all the corrections performed here. Our tests showed that precisions between 1 and 10 mas are expected from our algorithm.

We also applied our algorithm to the data cubes of HIP 049220, HIP 048480 and HIP 044245, which were observed in the \textit{K} band, with spaxel scales of 25, 100 and 250 mas, respectively. These data cubes are used to describe the methodologies in the next sections. As mentioned in Section 2, only one exposure for each one of these stars was retrieved. However, when there is more than one exposure per object, the best results are obtained when these exposures are combined in the form of a median, after the correction of the DAR.

\section{Spatial re-sampling}

\begin{figure*}
\begin{center}
  \includegraphics[scale=0.65]{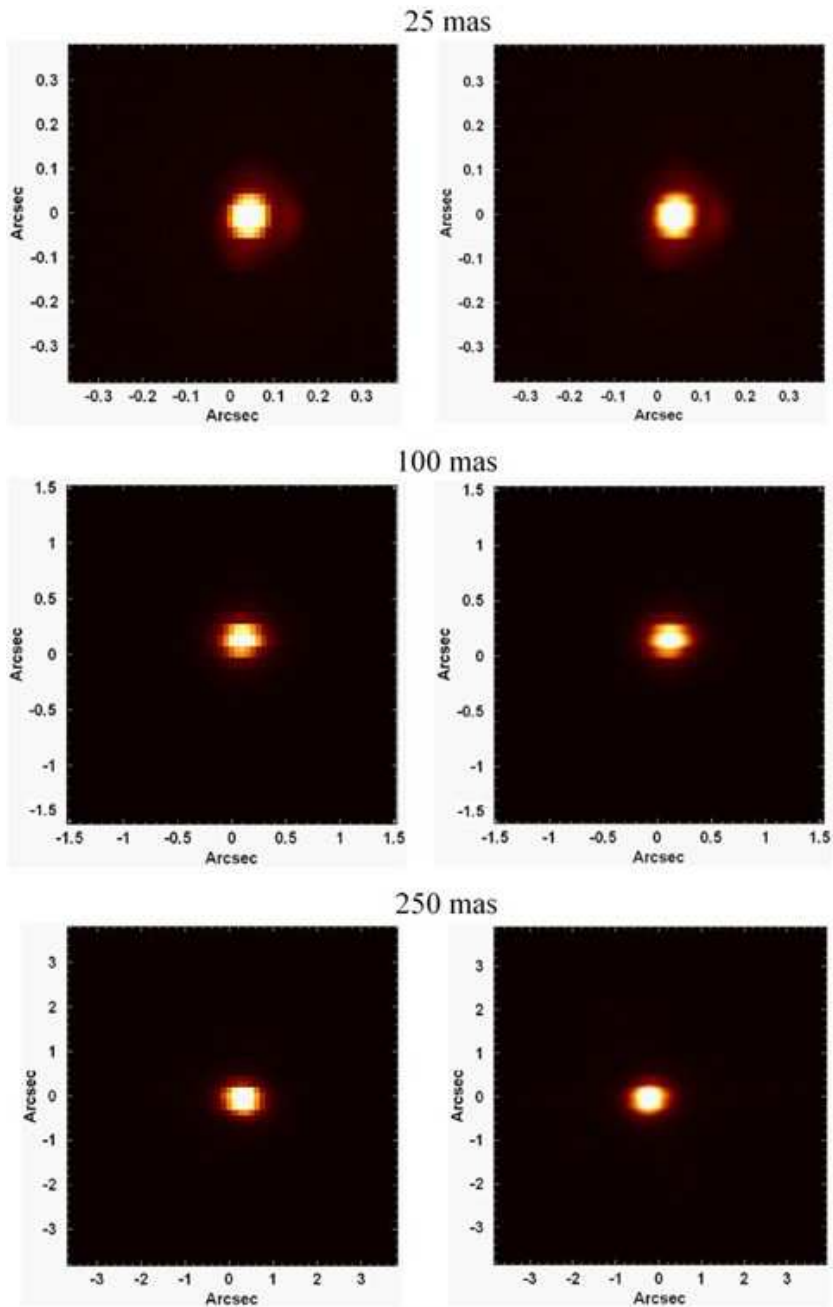}
  \caption{Images of intermediate-wavelength intervals of the data cubes of the stars (top) HIP 049220, (middle) HIP 048480 and (bottom) HIP 044245, (left) before and (right) after the spatial re-sampling.\label{fig6}}
\end{center}
\end{figure*}

\begin{figure*}
\begin{center}
  \includegraphics[scale=0.65]{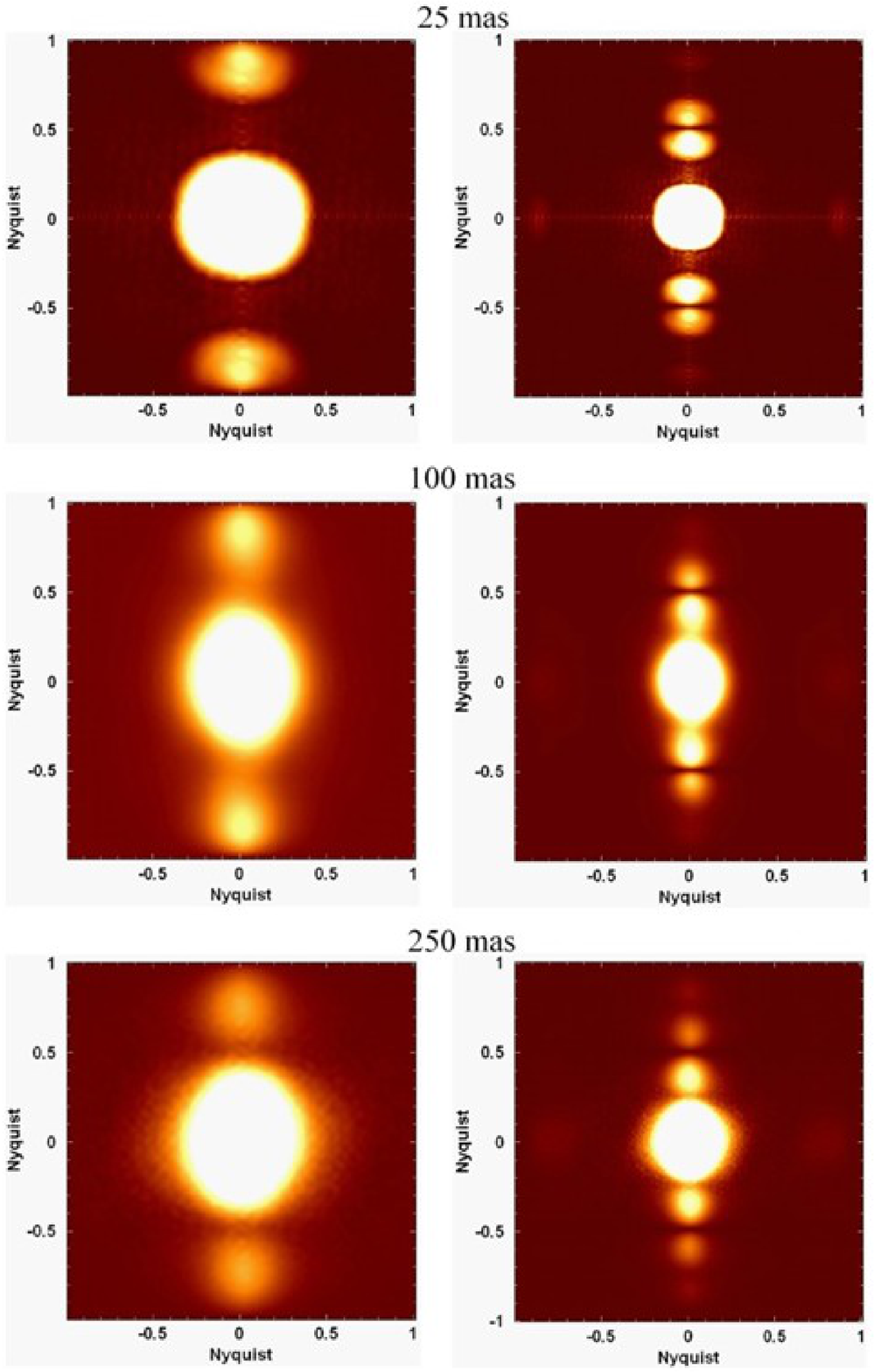}
  \caption{Moduli of the Fourier transforms of the images in Fig.~\ref{fig6}. Note that the high-frequency components introduced by the spatial re-sampling appear as bright regions farther from the centre of the Fourier transforms of the re-sampled data cubes.\label{fig7}}
\end{center}
\end{figure*}

The second step in our sequence for the treatment of SINFONI data cubes is the spatial re-sampling. We use an algorithm that reduces the size of the spaxels in the images of the data cubes, conserving the surface flux. Our algorithm is based on the task `frebin' of the {\sc idl} Astronomy User's Library\footnote{http://idlastro.gsfc.nasa.gov}. However, other tasks may also be used, as long as the surface flux is conserved. After the spatial re-sampling, we perform an interpolation of the values in each one of the lines and columns of the re-sampled images. Besides providing a better visualization of the contours of the structures in the images, the main improvement achieved with this procedure is related to the final spatial resolution, which is higher when the Richardson-Lucy deconvolution (the last step in our sequence - Section 7) is applied to re-sampled images. It should be noted that the spatial re-sampling of an image does not change its spatial resolution, but only improves its appearance.

In Paper I, we give details about the necessity of the use of an interpolation after the re-sampling of an image and about the fact that better results are obtained by applying the spatial re-sampling procedure, instead of simply reducing the data cubes with smaller spaxels. Since this previous discussion is entirely analogous to SINFONI data cubes, it will not be repeated here.

We usually apply the spatial re-sampling (followed by an interpolation) to SINFONI data cubes (observed with any fore-optics) in order to double the number of spaxels along the horizontal and vertical axis. Therefore, if data cubes observed with the fore-optics with scales of 25, 100 and 250 mas are reduced with spaxels with sizes of 12.5, 50 and 125 mas, respectively, then, the final sizes of their spaxels, after the spatial re-sampling, are 6.25, 25 and 62.5 mas, respectively. Fig.~\ref{fig6} shows images of intermediate-wavelength intervals of the data cubes with spaxel scales of 25, 100 and 250 mas, before and after the spatial re-sampling, following the strategy mentioned above.

In Fig,~\ref{fig6}, we can see that the spatial re-sampling improved the appearance of the images of the data cubes. However, we can also see the existence of some high spatial-frequency components, which appear as horizontal stripes, in the images of the re-sampled data cubes. A part of these high-frequency structures may have an instrumental origin and, therefore, was possibly already present in the data cubes before the spatial re-sampling. However, it is probable that some of them were introduced by the spatial re-sampling procedure. It is known that a spatial re-sampling introduces high spatial-frequency components in the images and that could explain, at least partially, the horizontal stripes detected in the images of the re-sampled data cubes. This side effect of the spatial re-sampling process is not problematic, as the high spatial-frequency structures are efficiently eliminated by the Butterworth spatial filtering (Section 5).  

\begin{figure*}
\begin{center}
  \includegraphics[scale=0.65]{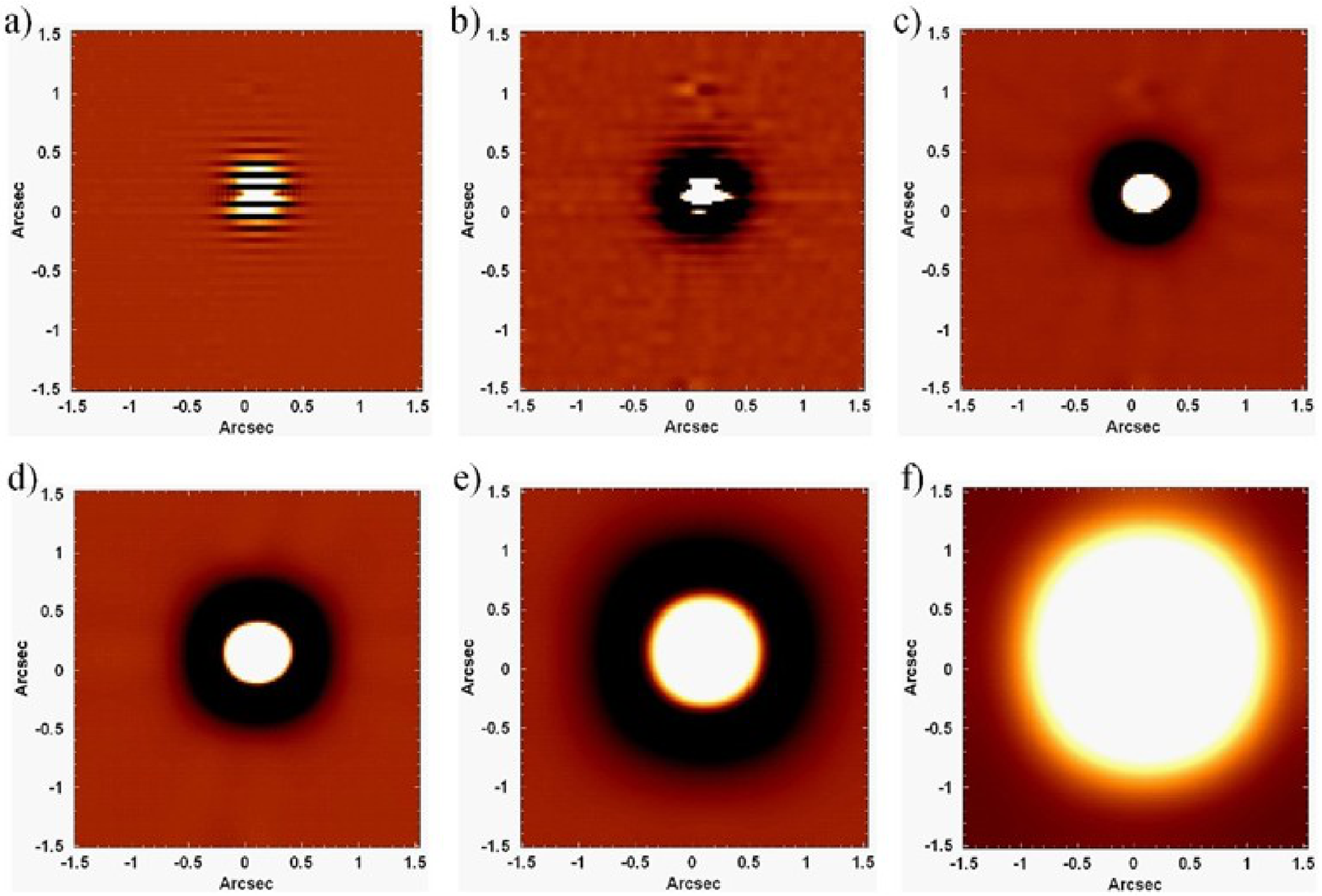}
  \caption{Images of the wavelet data cubes (a) $W_0$, (b) $W_1$, (c) $W_2$, (d) $W_3$, (e) $W_4$ and (f) $W_C$ of HIP 048480 (with spaxel scale of 100 mas), in an intermediate-wavelength interval.\label{fig8}}
\end{center}
\end{figure*}

To analyse in further detail the high-frequency components introduced by the spatial re-sampling, we calculated the Fourier transforms of all the images in Fig.~\ref{fig6}. The moduli of the obtained Fourier transforms are shown in Fig.~\ref{fig7}. The bright elliptical regions farther from the centre of the Fourier transforms represent the high spatial-frequency components introduced by the spatial re-sampling. The definition of the Nyquist frequency we are using is based on the spaxel sampling of each image. One interesting aspect of Fig.~\ref{fig7} is that the high-frequency components of the images are much more intense along the vertical axis than along the horizontal axis. This characteristic can also be seen in the re-sampled images in Fig.~\ref{fig6}, as the high-frequency components appear mainly as horizontal stripes and not vertical stripes in the images. This behaviour is very common in SINFONI data cubes and represents a difference from what we have observed in NIFS data cubes (Paper I), in which the high-frequency components along the horizontal and vertical axis have similar intensities. Another important feature of the Fourier transforms in Fig.~\ref{fig7} is that the region corresponding to the low spatial frequencies has an elliptical shape. Again, this is characteristic of SINFONI data cubes and represents a difference from NIFS data cubes, in which the low-frequency region in the Fourier transforms has a shape similar to the product of an ellipse by a rectangle.   

Another possible side effect of the spatial re-sampling of the images in a data cube is the degradation of the signal-to-noise (S/N) ratio of the spectra. However, in our previous experiences \citep{men12}, we have never observed measurable considerable degradations. Therefore, considering the benefits provided by the spatial re-sampling and the non-significant side effects, we conclude that the spatial re-sampling of SINFONI data cubes is recommended.

\section{Butterworth spatial filtering}

\begin{figure*}
\begin{center}
  \includegraphics[scale=0.65]{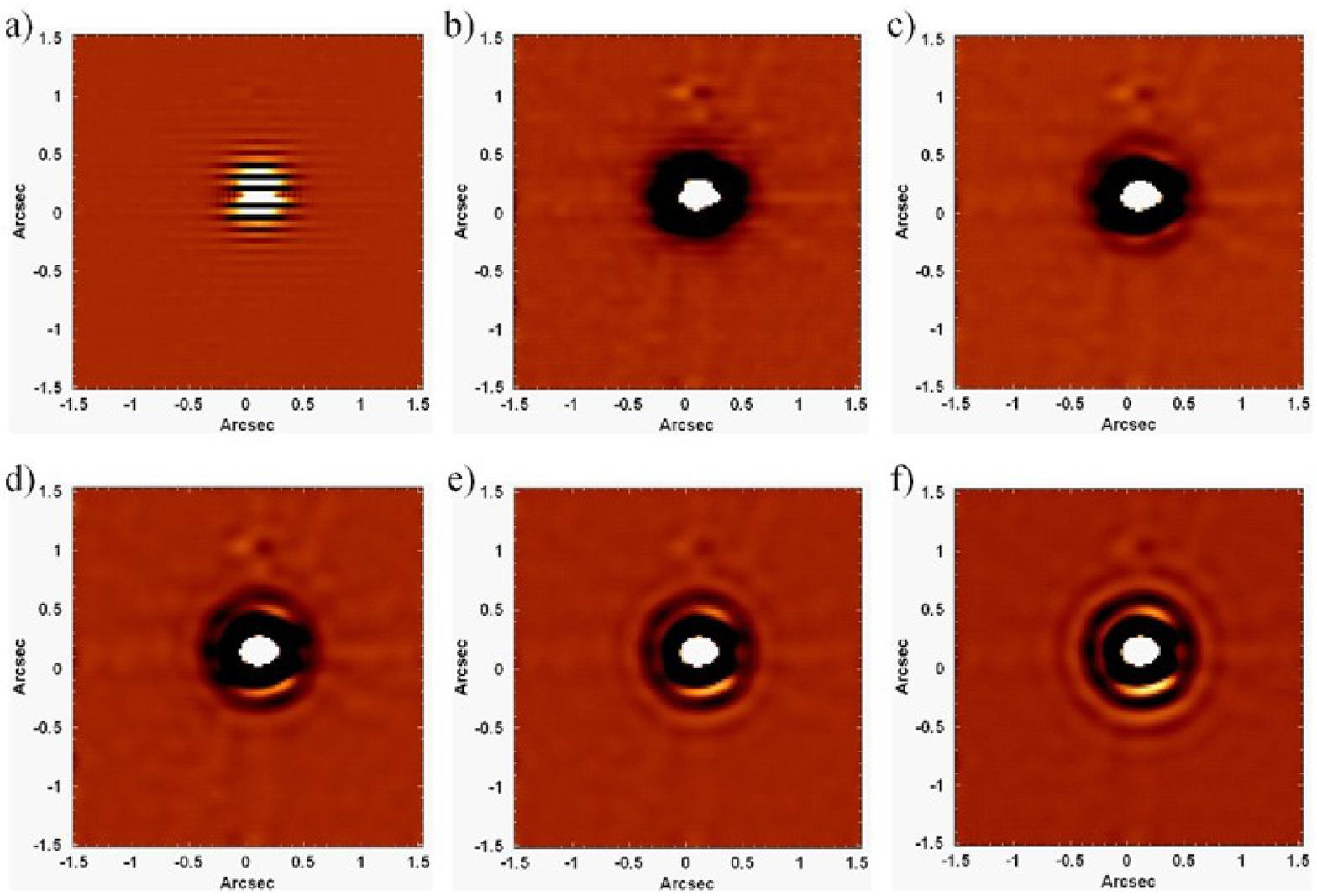}
  \caption{Images of an intermediate-wavelength interval of (a) the non-filtered $W_0$ data cube of HIP 048480 (with spaxel scale of 100 mas) and of the corresponding filtered data cubes, using (b) $n = 2$, (c) $n = 3$, (d) $n = 4$, (e) $n = 5$ and (f) $n = 6$.\label{fig9}}
\end{center}
\end{figure*}

The third step in our sequence for the treatment of SINFONI data cubes is the Butterworth spatial filtering \citep{gon02}, which is performed in the frequency domain. This procedure is described in detail in Paper I. We use a script that applies the filtering to all the images of the data cube, in order to remove the high spatial-frequency noise.

\begin{figure*}
\begin{center}
  \includegraphics[scale=0.65]{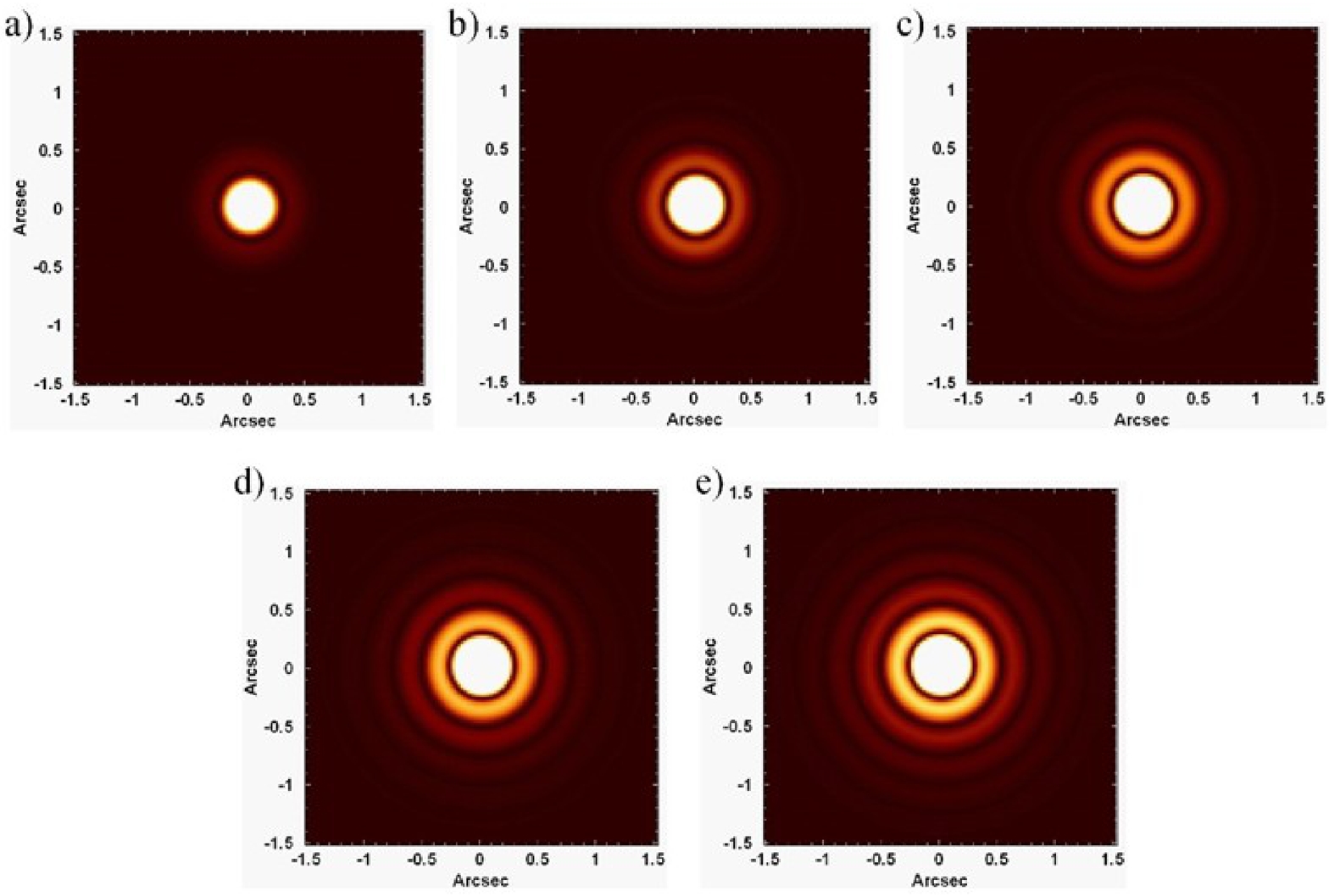}
  \caption{Moduli of the inverse Fourier transforms of the circular Butterworth filters, given by equation (1) (with $a = b$), with orders (a) $n = 2$, (b) $n = 3$, (c) $n = 4$, (d) $n = 5$ and (e) $n = 6$. All the filters were obtained using a cut-off frequency of 0.21 Ny.\label{fig10}}
\end{center}
\end{figure*}

The general formula of a low-pass Butterworth filter of order $n$ is given by equation (7) of Paper I. Since the low-frequency region of the Fourier transforms of SINFONI data cubes have typical elliptical shapes (see Fig.~\ref{fig7}), the best choice for the Butterworth filter is an elliptical one, which is given by

\begin{equation}
H\left(u,v\right) = \left\{\frac{1}{1 + \left[\sqrt{\left(\frac{u - u_0}{a}\right)^2 + \left(\frac{v - v_0}{b}\right)^2}\right]^{2n}}\right\},
\end{equation}

In the previous equation, the point $(u_0,v_0)$ is located at the centre of the Fourier transform, $n$ is the order of the filtering, $a$ is the cut-off frequency along the horizontal axis and $b$ is the cut-off frequency along the vertical axis. Our tests revealed that, considering the eccentricities of the Butterworth filters involved, it is very difficult to detect measurable differences between the results obtained using an elliptical filter and using a simple circular filter (which is equivalent to assume $a = b$ in equation 1). Therefore, we use a circular filter throughout this paper.

Fig.~\ref{fig7} indicates that a cut-off frequency of approximately 0.21 Ny seems to be adequate for the Butterworth spatial filtering of the data cube of the standard star observed with spaxel scale of 100 mas. Following the same approach we used in Paper I, we applied a spatial wavelet decomposition \citep{sta06}, using the \`A Trous algorithm, to each image of this data cube. For more details about the wavelet decomposition of data cubes and about the \`A Trous algorithm, see Paper I. Fig.~\ref{fig8} shows images of an intermediate-wavelength interval of the obtained wavelet data cubes $W_0$, $W_1$, $W_2$, $W_3$, $W_4$ and $W_C$. The high spatial-frequency components appear very clearly in the image of $W_0$ as horizontal stripes. The interference of these structures in the image is so significant that the star cannot be seen directly. Part of these high-frequency components was probably introduced by the spatial re-sampling, although it is also possible that at least some of these stripes have an instrumental origin (as discussed in Section 4). In fact, the horizontal stripes were also easily detected in the image of an intermediate-wavelength interval of this data cube, after the spatial re-sampling (see Fig.~\ref{fig6}). The high spatial-frequency structures are much less abundant in the image of $W_1$ and are practically absent in the other images in Fig.~\ref{fig8}.

We then applied the Butterworth spatial filtering to the $W_0$ data cube of the standard star observed with spaxel scale of 100 mas, using a filter given by equation (1) with a cut-off frequency of 0.21 Ny and with $n = 2$, $n = 3$, $n = 4$, $n = 5$ and $n = 6$. Fig.~\ref{fig9} shows images of an intermediate-wavelength interval of $W_0$, filtered with these values of $n$.

\begin{figure*}
\begin{center}
  \includegraphics[scale=0.65]{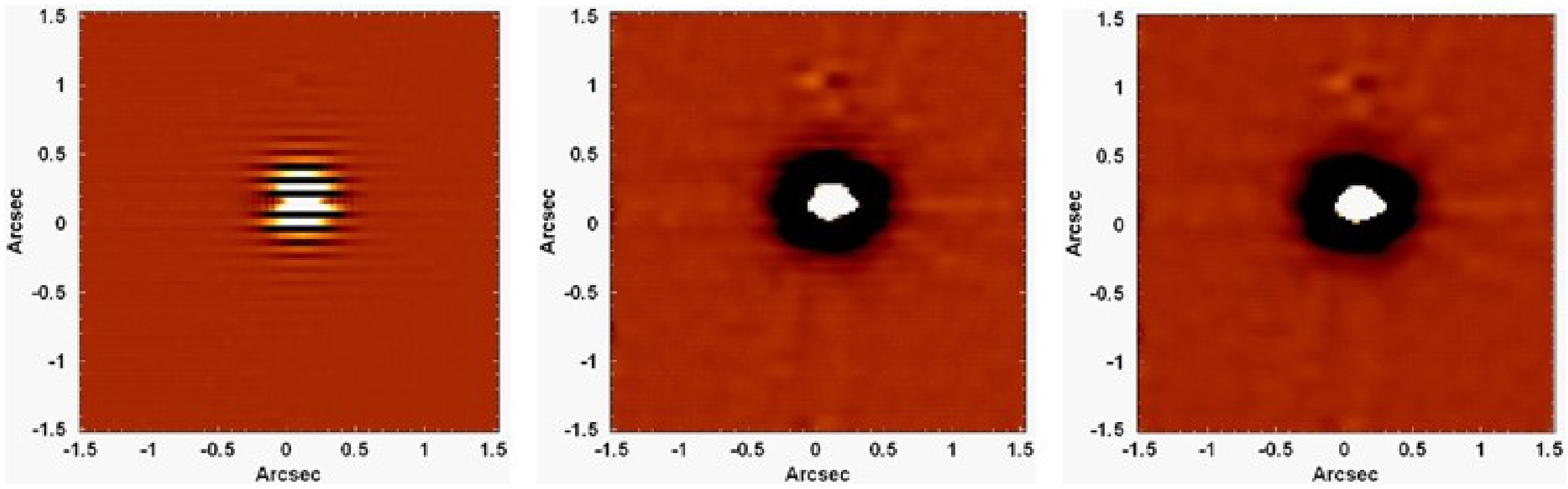}
  \caption{Images of an intermediate-wavelength interval of the $W_0$ data cube of HIP 048480 (with spaxel scale of 100 mas) (left) before the Butterworth spatial filtering, (centre) after the Butterworth spatial filtering using a simple circular filter with $n = 2$ and a cut-off frequency of 0.21 Ny and (right) after the Butterworth spatial filtering using a squared circular filter with $n = 2$ and cut-off frequency of 0.26 Ny. Note that some traces of the high-frequency noise are still visible in the image from the data cube filtered with a simple circular filter, but are absent in the image from the data cube filtered with a squared circular filter.\label{fig11}}
\end{center}
\end{figure*}

A simple analysis of Fig.~\ref{fig9} indicates that all the filtering procedures applied to $W_0$, with different values of $n$, removed a substantial amount of high spatial-frequency structures and, as a consequence, the star became visible in all filtered images of $W_0$. However, the images of the data cubes filtered with orders higher than 2 show the existence of `concentric rings' around the central star. It is also easy to see that the higher the value of $n$, the more intense and more numerous are the rings. As discussed in Paper I, this behaviour can be explained by the fact that, based on the convolution theorem, a Butterworth spatial filtering of an image is equivalent to the convolution of the image with the inverse Fourier transform of the Butterworth filter. Considering this and the aspect of the inverse Fourier transforms of the Butterworth filters with different values of $n$ (shown in Fig.~\ref{fig10}), the reason for the existence of concentric rings in the images of data cubes filtered with orders higher than 2 becomes very clear.

There are some points that are worth mentioning here. First, although the concentric rings introduced by the Butterworth spatial filtering with high orders were evident in Fig.~\ref{fig9}, it is much more difficult to detect them in images of original SINFONI data cubes (without a wavelet decomposition), because they are usually immersed in the PSFs of the observations. They were so clearly seen here because we used the $W_0$ data cube in this discussion. Besides that, considering the way these rings are introduced in the images, they are only detected around point-like sources. Finally, in a Butterworth spatial filtering, there is an issue associated with the periodicity of Fourier transforms. Such periodicity is a consequence of the way a Fourier transform is defined. The results obtained by a convolution between two functions with and without periodicity are different. Therefore, since the Butterworth spatial filtering of an image is equivalent to the convolution between the image and the inverse Fourier transform of the filter, we conclude that this process is affected by the periodicity of the Fourier transforms. To avoid this problem, we apply a `padding' procedure, which will not be discussed here, but a detailed description can be found in \citet{gon02}.

\begin{figure*}
\begin{center}
  \includegraphics[scale=0.66]{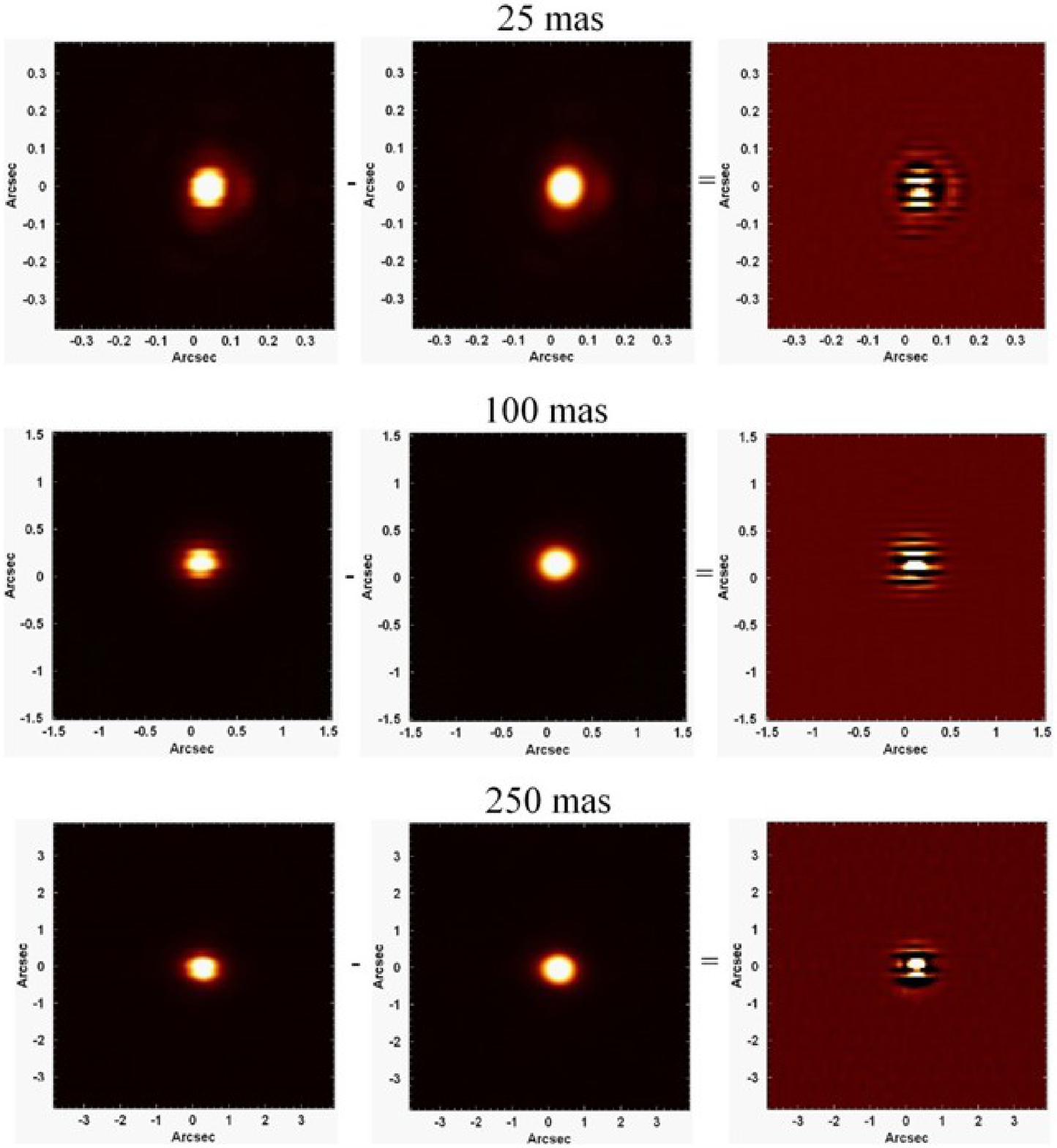}
  \caption{Images of intermediate-wavelength intervals of the data cubes of (top) HIP 049220, (middle) HIP 048480 and (bottom) HIP 044245. The images at left represent the data cubes before the Butterworth spatial filtering. The images at centre represent the data cubes after the Butterworth spatial filtering with cut-off frequencies of (top) 0.29 Ny, (middle) 0.26 Ny and (bottom) 0.27 Ny. The images at right represent the data cubes corresponding to the difference between the non-filtered and filtered ones.\label{fig12}}
\end{center}
\end{figure*} 

\begin{figure*}
\begin{center}
  \includegraphics[scale=0.55]{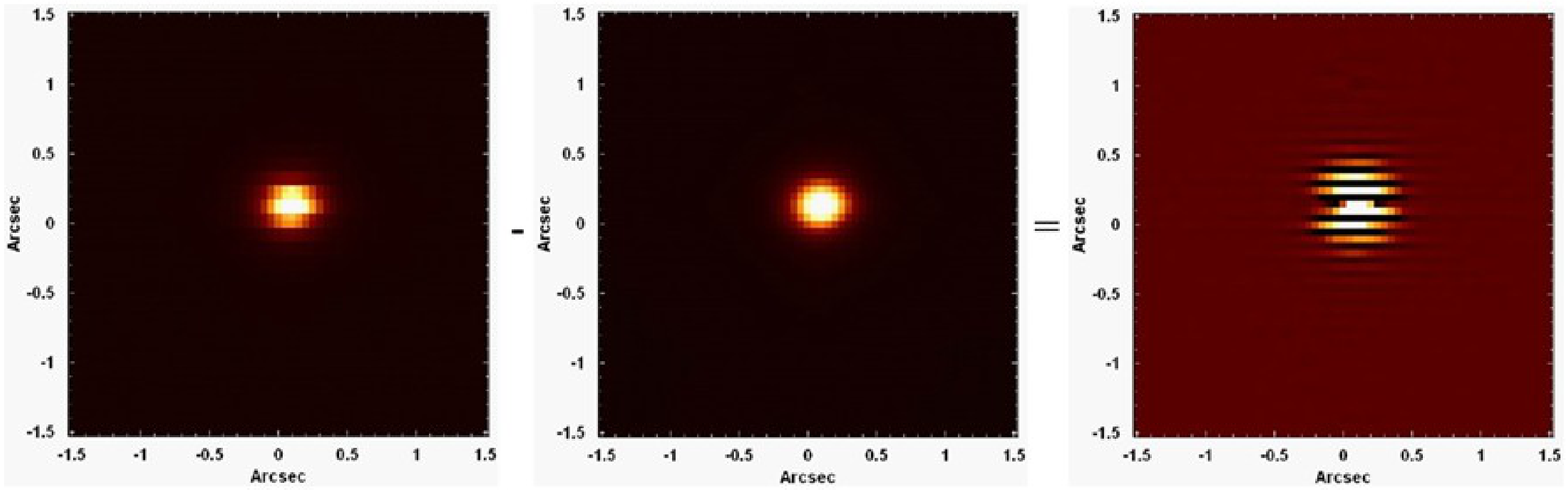}
  \caption{Left: image of an intermediate-wavelength interval of the non-filtered data cube of HIP 048480 (with spaxel scale of 100 mas), with spaxels of 50 mas. Centre: image of the same intermediate-wavelength interval of the filtered data cube of HIP 048480, with spaxels of 50 mas. The filtering was applied using a squared circular filter with $n = 2$ and a cut-off frequency of 0.55 Ny. Right: image of the same intermediate-wavelength interval of the data cube corresponding to the difference between the previous two.\label{fig13}}
\end{center}
\end{figure*}

Since the Butterworth spatial filtering with $n = 2$ removed a substantial amount of high spatial-frequency noise and did not introduce visible artefacts in the images, we can say that this seems to be the most adequate value of $n$ to be used in the Butterworth spatial filtering of SINFONI data cubes. However, a careful analysis of Fig.~\ref{fig9} reveals that, although the Butterworth spatial filtering of $W_0$ with $n = 2$ has removed most of the high spatial-frequency noise, some traces of this noise are still visible in the image of the filtered data cube. This suggests that this filtering could be improved. One possible strategy is to use a Butterworth filter, with $n = 2$, given by the following equation:

\begin{equation}
H\left(u,v\right) = \left\{\frac{1}{1 + \left[\sqrt{\left(\frac{u - u_0}{a}\right)^2 + \left(\frac{v - v_0}{b}\right)^2}\right]^{2n}}\right\}^2,
\end{equation}
which corresponds to the product between two identical elliptical filters. The use of the Butterworth filter given by equation (2) is equivalent to apply two successive Butterworth filterings with the same elliptical filter. If this process is performed with $n = 2$, no significant ring will be introduced in the filtered images, but the efficacy of the filtering will be improved. Again, we assume $a = b$ in equation (2), which is equivalent to use a squared circular filter. It is important to mention, however, that the cut-off frequency of a squared circular filter has to be a little higher than the cut-off frequency of a simple circular filter with the same value of $n$, in order to not compromise the PSF of the observation. In the case of the data cube of the standard star observed with spaxel scale of 100 mas, our tests revealed that the most appropriate value for the cut-off frequency of the squared circular filter is 0.26 Ny. Fig.~\ref{fig11} shows images of an intermediate-wavelength interval of the $W_0$ data cubes of this star before the Butterworth spatial filtering, after the Butterworth spatial filtering using a simple circular filter with $n = 2$ and cut-off frequency of 0.21 Ny, and after the Butterworth spatial filtering using a squared circular filter with $n = 2$ and cut-off frequency of 0.26 Ny.

Fig.~\ref{fig11} shows that the Butterworth spatial filtering using a squared circular filter with $n = 2$ removed a little more high spatial-frequency noise than the Butterworth spatial filtering using a simple circular filter with $n = 2$. Therefore, we conclude that the Butterworth spatial filtering using a squared circular filter with $n = 2$ is the most appropriate to be applied to SINFONI data cubes. Although the previous discussion was based on the results obtained with the data cube of the standard star observed with the fore-optics with spaxel scale of 100 mas, the conclusions are valid for observations taken with any fore-optics.

The seeing conditions of the observing night and the effect of the AO (which does not have the same efficiency in all observations) may affect the value of the cut-off frequency to be used in the Butterworth spatial filtering. In the case of the fore-optics with an FOV of 8.0 arcsec (in which the AO is frequently not used), the values of the cut-off frequencies will depend basically on the seeing conditions of the night (and also on the size of the spaxels of the instrument, which, as explained in Section 7, affect the PSF of the observation). In the case of the other two fore-optics, however, the cut-off frequencies will depend mainly on the efficiency of the AO (and, again, on the size of the spaxels of the instrument). In our previous experiences \citep{men12}, we have applied the Butterworth spatial filtering to SINFONI data cubes using cut-off frequencies between 0.20 and 0.35 Ny. 

Using a filter given by equation (2) and $n = 2$, we applied the Butterworth spatial filtering to the original data cubes (without any wavelet decomposition) of the standard stars observed with spaxel scales of 25, 100 and 250 mas, with cut-off frequencies of 0.29, 0.26 and 0.27 Ny, respectively. Fig.~\ref{fig12} shows images of intermediate-wavelength intervals of these data cubes, before and after the Butterworth spatial filtering, and also of the data cubes corresponding to the difference between the filtered and non-filtered data cubes of these objects. We can see that most of the high spatial-frequency noise (in the form of horizontal stripes) was removed from the images of the data cubes by the filtering procedure, without the introduction of detectable artefacts.

One important topic to be mentioned here is about the origin of the high spatial-frequency components in the images of the data cubes with spaxel scales of 25, 100 and 250 mas. As discussed in Paper I, such high-frequency components are not introduced by the interpolation we apply after the spatial re-sampling. We applied a Butterworth spatial filtering to the non re-sampled data cube of the star observed with spaxel scale of 100 mas (using a squared circular filter, with n=2), in order to determine if these high-frequency components were completely introduced by the spatial re-sampling or were, at least some of them, present in the non re-sampled data cubes. In this case, we verified that the most appropriate value for the cut-off frequency to be used in the Butterworth spatial filtering is 0.55 Ny. Fig.~\ref{fig13} shows images of an intermediate-wavelength interval of the non re-sampled data cube of this star, before and after the Butterworth spatial filtering, and also of the data cube corresponding to the difference between the filtered and non-filtered ones.

Fig.~\ref{fig13} reveals that the non re-sampled data cube of the star observed with spaxel scale of 100 mas has essentially the same high spatial-frequency components observed in the re-sampled data cube (Fig.~\ref{fig12}). Therefore, we conclude that, although the spatial re-sampling procedure has certainly introduced high spatial-frequency components in the images of this data cube (as the Fourier transforms in Fig.~\ref{fig7} show), most of them were already present in the original data cube. This discussion was based, only, on the data cube with spaxel scale of 100 mas; however, the results obtained with data cubes in the other fore-optics are entirely analogous.  

\begin{figure*}
\begin{center}
  \includegraphics[scale=0.52]{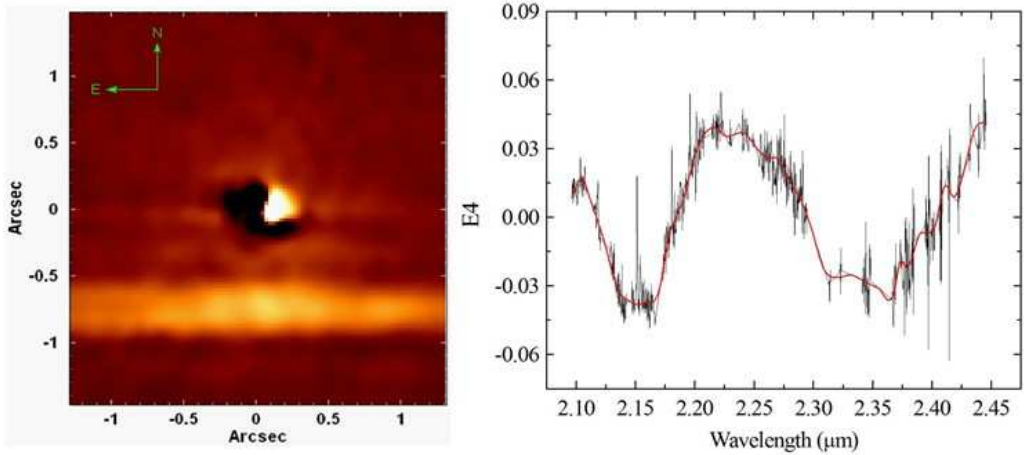}
  \caption{Tomogram and eigenspectrum corresponding to eigenvector \textit{E}4, obtained with PCA Tomography of the data cube of Centaurus A, with spaxel scale of 100 mas, after the removal of the spectral lines. The spline fitted to the eigenspectrum is shown in red. This eigenvector is dominated by the instrumental fingerprint, but the AGN emission can be detected in the tomogram. However, the effect of this emission is of second order.
\label{fig14}}
\end{center}
\end{figure*}

\begin{figure*}
\begin{center}
  \includegraphics[scale=0.52]{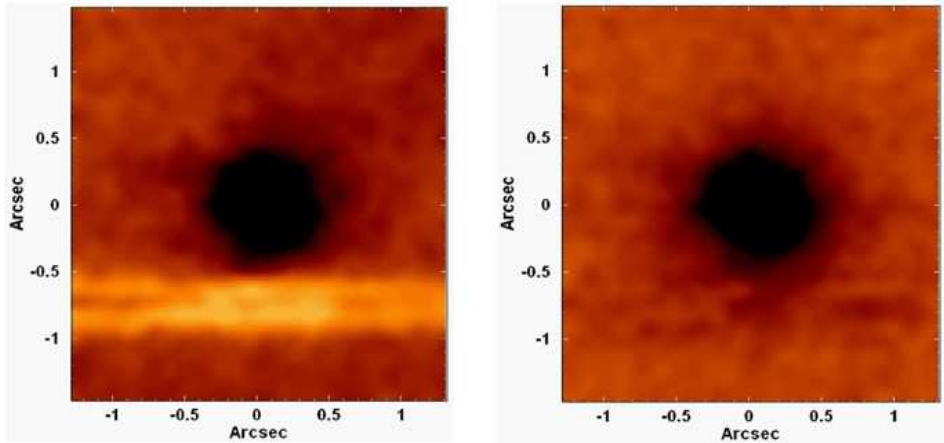}
  \caption{Results of the subtraction of the image corresponding to the wavelength interval 2.1481 - 2.1530 $\mu$m from the image corresponding to the wavelength interval 2.0988 - 2.1037 $\mu$m of the data cube of Centaurus A, with spaxel scale of 100 mas. The image at left was obtained before the instrumental fingerprint removal and the image at right was obtained after the fingerprint removal.\label{fig15}}
\end{center}
\end{figure*}

\begin{figure*}
\begin{center}
  \includegraphics[scale=0.65]{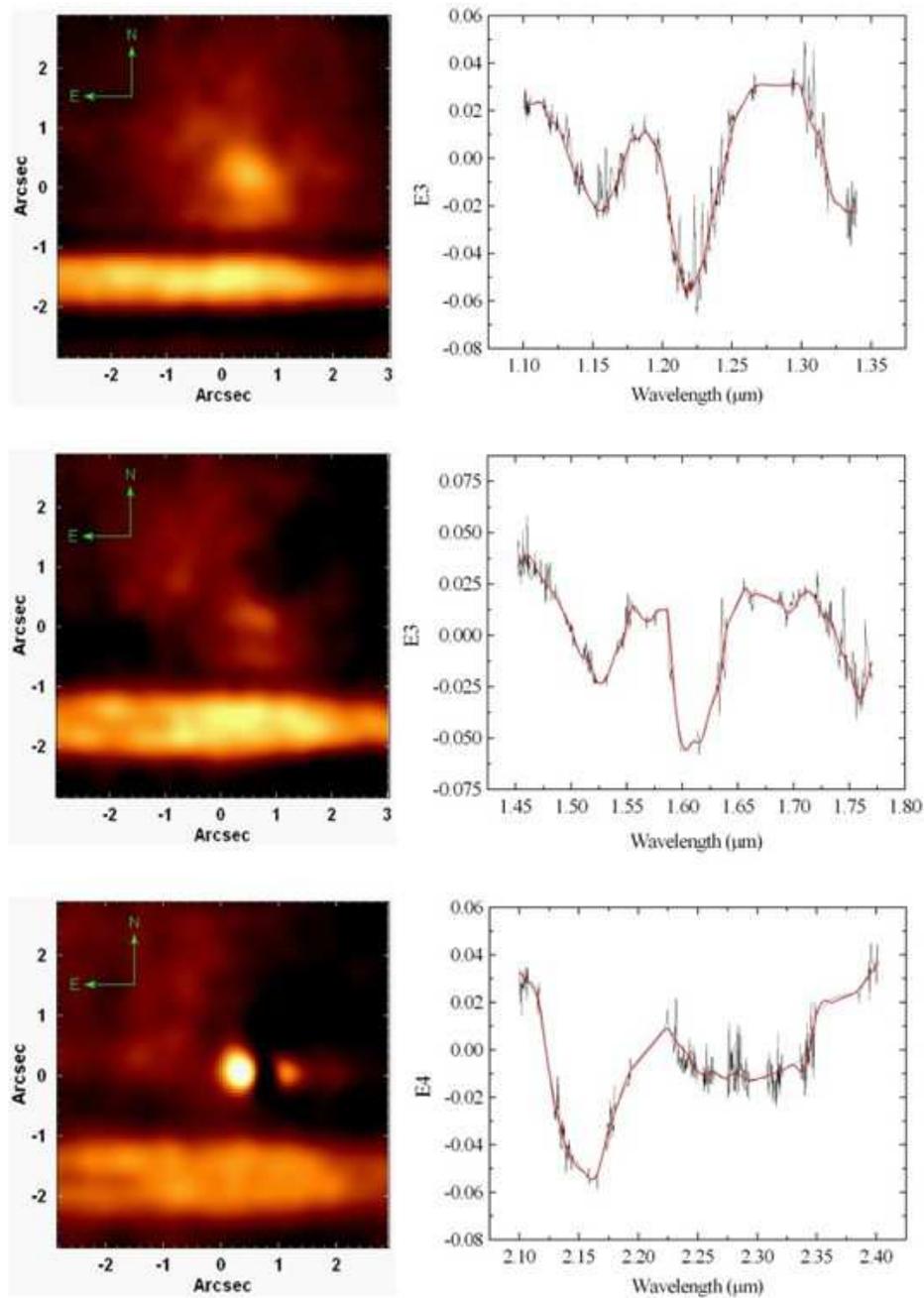}
  \caption{Tomograms and eigenspectra corresponding to the eigenvectors related to the instrumental fingerprints in the data cubes of Centaurus A, with spaxel scale of 250 mas, after the removal of the spectral lines, in the (top) \textit{J}, (middle) \textit{H} and (bottom) \textit{K} bands. The splines fitted to the eigenspectra are shown in red. Although the eigenvectors are dominated by the instrumental fingerprint, the AGN emission can be seen in the tomogram obtained from the data cube in the \textit{K} band and the shape of the diffuse stellar component is visible in all tomograms. However, as mentioned for Fig.~\ref{fig14}, the effects of this emission from the AGN and from the stellar component are of second order.\label{fig16}}
\end{center}
\end{figure*}

\begin{figure*}
\begin{center}
  \includegraphics[scale=0.65]{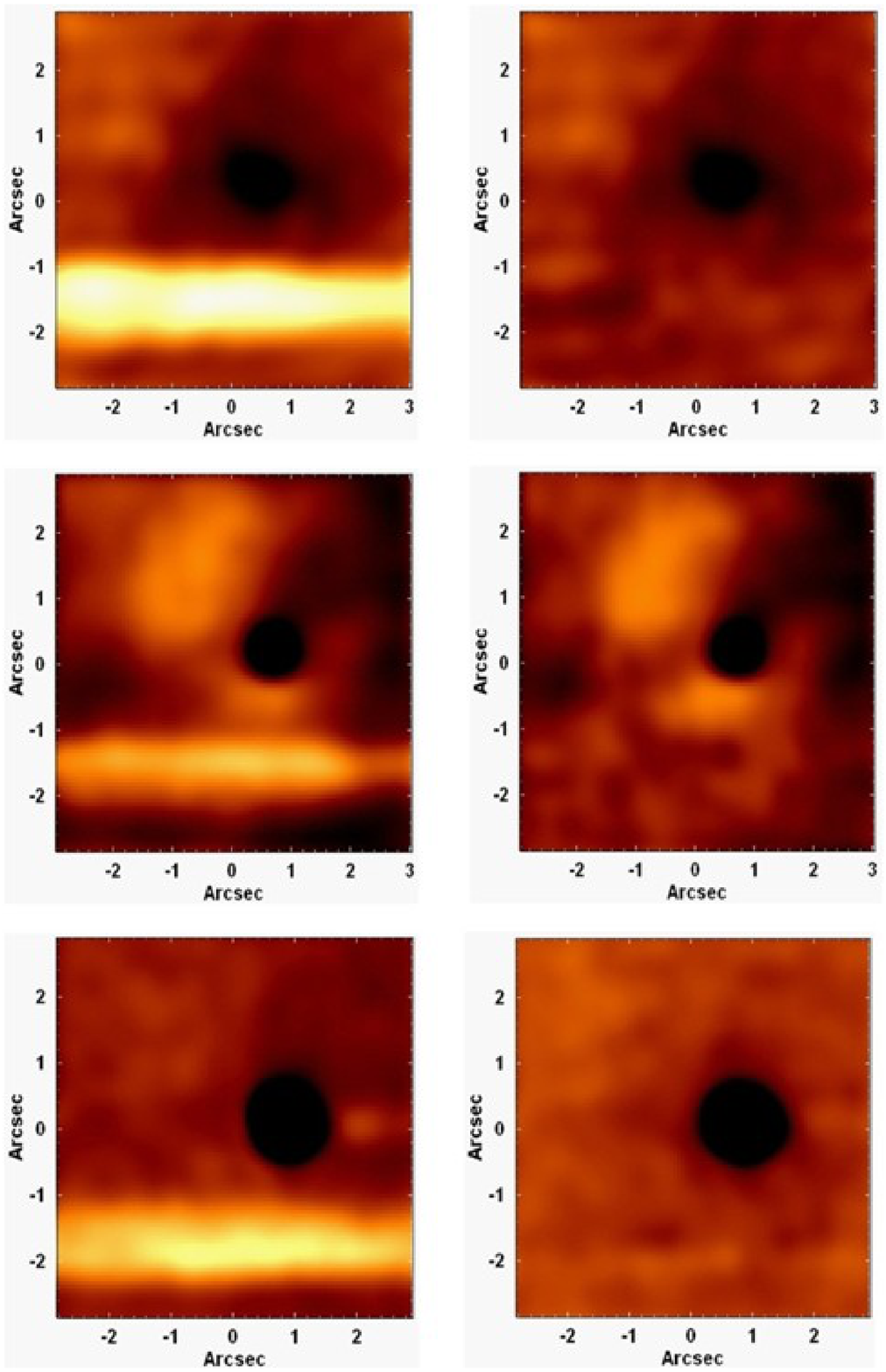}
  \caption{Top: results of the subtraction of the image corresponding to the wavelength interval 1.2091 - 1.2156 $\mu$m from the image corresponding to the wavelength interval 1.1727 - 1.1792 $\mu$m (left) before and (right) after the instrumental fingerprint removal of the data cube of Centaurus A. Middle: results of the subtraction of the image corresponding to the wavelength interval 1.5905 - 1.6007 $\mu$m from the image corresponding to the wavelength interval 1.5480 - 1.5582 $\mu$m (left) before and (right) after the instrumental fingerprint removal of the data cube of Centaurus A. Bottom: results of the subtraction of the image corresponding to the wavelength interval 2.1368 - 2.1454 $\mu$m from the image corresponding to the wavelength interval 2.1098 - 2.1184 $\mu$m (left) before and (right) after the instrumental fingerprint removal of the data cube of Centaurus A. All data cubes were obtained with the fore-optics with spaxel scale of 250 mas.\label{fig17}}
\end{center}
\end{figure*}

\section{`Instrumental fingerprint' removal}

After the Butterworth spatial filtering, the next step in our data treatment is the `instrumental fingerprint' removal. Unlike the NIFS instrumental fingerprint we described in Paper I, which is relatively subtle and may not be relevant for most studies, the SINFONI instrumental fingerprint is significantly more intense and may affect many studies. This feature usually takes the form of a large horizontal stripe at the bottom of the images and shows a characteristic low-frequency spectral signature. One example of a study affected by the SINFONI instrumental fingerprint is the work made by \citet{neu07} of data cubes of the nuclear region of the Centaurus A galaxy. These data were obtained in the \textit{H} and \textit{K} bands, using the fore-optics with an FOV of 3.2 arcsec. The [Fe II] images in that study were significantly affected by the fingerprint. The process for removing this feature is the same we use for removing instrumental fingerprints of NIFS data cubes (see Paper I) and is based on the PCA Tomography technique \citep{ste09}.

PCA Tomography involves the use of the Principal Component Analysis (PCA - Murtagh \& Heck 1987; Fukunaga 1990) technique on data cubes. In this application of PCA, the spectral pixels of the data cube are taken as the variables of a coordinate system and the spaxels are taken as the observables. PCA then creates a new system of uncorrelated coordinates (called eigenvectors), which are linear combinations of the original coordinates (the wavelengths) and are orthogonal to each other. These eigenvectors are ordered in a way that eigenvector \textit{E}1 is associated with the highest fraction of the data variance, eigenvector \textit{E}2 is associated with the second highest fraction of the data variance and so on. Due to the way the eigenvectors are defined in the case of PCA Tomography, we call them eigenspectra. The projections of the observables (spaxels) on the eigenvectors are images, which we call tomograms. 

The process for removing the instrumental fingerprint of SINFONI data cubes involves the following steps:

\begin{itemize}

\item removal of the spectral lines (both emission and absorption lines) of the data cube,

\item application of PCA Tomography to the data cube without emission lines,

\item selection of the eigenvetors related to the instrumental fingerprint (the eigenvectors related to the instrumental fingerprint must have the spectral signature of the fingerprint and the corresponding tomograms must have the spatial morphology of a large horizontal stripe at the bottom of the images),

\item fit of splines to the eigenspectra related to the instrumental fingerprint,

\item construction of a data cube using the splines fitted to the eigenspectra related to the instrumental fingerprint and the corresponding tomograms (the resulting data cube will contain only the fingerprint),

\item subtraction of the data cube obtained above from the original one.

\end{itemize}

The discussion of all previous steps can be found in Paper I. In order to give an example of the efficacy of our method for instrumental fingerprint removal on SINFONI data cubes, we used a data cube of the nuclear region of Centaurus A, which was observed with SINFONI, in the \textit{K} band, using the fore-optics with an FOV of 3.2 arcsec. This is one of the data cubes analysed by \citet{neu07} and it was treated with all the techniques described before. After the removal of the spectral lines, we applied the PCA Tomography on the resulting data cube. After a careful analysis, we verified that the obtained eigenvector \textit{E}4 reveals very clearly the instrumental fingerprint typically found in SINFONI data cubes. The eigenspectrum and the tomogram corresponding to eigenvector \textit{E}4 are shown in Fig.~\ref{fig14}. The spline fitted to the eigenspectrum is shown in red. Eigenvector \textit{E}4 explains $\sim 7.0 \times 10^{-4}$ per cent of the variance of the data cube obtained after the removal of the spectral lines.

Eigenspectrum \textit{E}4 in Fig.~\ref{fig14}, and the fitted spline, reveal very clearly the spectral signature of the instrumental fingerprint. This spectral signature shows two main broad features: one around 2.15 $\mu$m and the other around 2.35 $\mu$m. The feature at 2.15 $\mu$m is the most characteristic of the instrumental fingerprint of SINFONI data cubes obtained (with any fore-optics) in the \textit{K} band. The feature at 2.35 $\mu$m may also appear, although it is more diffuse in some cases. The small fraction of the data variance explained by eigenvector \textit{E}4 suggests that this instrumental fingerprint is not intense enough to compromise any analysis or measurement. However, as mentioned before, this artefact may affect considerably the images of the data cubes \citep{neu07}. In order to illustrate the effect of this fingerprint on a scientific analysis, we constructed images of the wavelength intervals 2.0988 - 2.1037 $\mu$m and 2.1481 - 2.1530 $\mu$m of the data cube of Centaurus A. The second image was then subtracted from the first one. This procedure was applied to the data cube before and after the instrumental fingerprint removal. The obtained images, shown in Fig.~\ref{fig15}, reveal the considerable improvement provided by our correction. The instrumental fingerprint can be easily seen in the non-corrected image, but essentially no fingerprint is detected in the corrected image. Since the fingerprint is associated with the spectral continuum and not with the emission lines, analysis involving the continuum are usually more affected than other types of analysis. One important study of the spectral continuum that could be affected by the instrumental fingerprint is the spectral synthesis. However, as mentioned before, in some cases, even studies involving the images of emission lines may be affected by the fingerprint \citep{neu07}.

In order to compare the observed instrumental fingerprint in different spectral bands, we also used data cubes of the nuclear region of Centaurus A, observed, in the \textit{J}, \textit{H} and \textit{K} bands, using the fore-optics with an FOV of 8.0 arcsec. We chose this specific fore-optics because it was the only one with available observations in the three spectral bands of this object. We followed the procedure described above and applied PCA Tomography to these three data cubes, after the removal of the spectral lines. Fig.~\ref{fig16} shows the obtained eigenspectra and tomograms associated with the fingerprint in the data cubes. The splines fitted to the eigenspectra are shown in red. Eigenvectors \textit{E}3 (in the \textit{J} band), \textit{E}3 (in the \textit{H} band) and \textit{E}4 (in the \textit{K} band) explain $\sim 0.142$, $\sim 0.075$ and $\sim 0.003$ percent, respectively, of the data variances of the corresponding data cubes, after the removal of the spectral lines. We can see that the spatial morphology and the spectral signature of the fingerprint in the three spectral bands are actually very similar. The fingerprint has the appearance of a large horizontal stripe at the bottom of the images and the two broad features of the spectral signature are located around 1.15 and 1.22 $\mu$m, in the \textit{J} band, around 1.52 and 1.60 $\mu$m, in the \textit{H} band, and, as mentioned before, around 2.15 and 2.30 $\mu$m, in the \textit{K} band. As Figs. 14 and 16 reveal, the characteristics of the instrumental fingerprint do not change with the fore-optics used in the observation. Again, in order to show the effect of the fingerprint on a scientific analysis of these data cubes, we constructed images of the wavelength intervals 1.1727 - 1.1792 $\mu$m and 1.2091 - 1.2156 $\mu$m, in the \textit{J} band, 1.5480 - 1.5582 $\mu$m and 1.5905 - 1.6007 $\mu$m, in the \textit{H} band, and 2.1098 - 2.1184 $\mu$m and 2.1368 - 2.1454 $\mu$m, in the \textit{K} band. Then, for each band, we subtracted the second wavelength interval from the first one. We applied this procedure before and after the instrumental fingerprint removal. Fig.~\ref{fig17} shows the results. It is easy to see the substantial improvements provided by our treatment procedure.

The spectral characteristics of the eigenspectra in Figs. 14 and 16, together with the spatial morphologies of the corresponding tomograms, clearly indicate that these eigenvectors are mainly associated with the instrumental fingerprint. However, in our previous experiences \citep{men12}, we verified that eigenvectors dominated by a given phenomenon may also show contaminations by other phenomena. The tomograms obtained from the data cubes of NGC 5643 in the \textit{K} band, shown in Figs 14 and 16, for example, have structures related to the active galactic nucleus (AGN) emission. The shape of the diffuse stellar component may also be seen in the tomograms from the data cubes in the \textit{J}, \textit{H} and \textit{K} bands, shown in Fig.~\ref{fig16}. However, the effects of this emission from the AGN and from the stellar component are of second order, as the subtraction of the data cubes constructed using the eigenvectors associated with the fingerprint from the original data cubes removed the fingerprint but did not result in measurable changes in the observed emission. In other words, we can say that the eigenvectors in Figs 14 and 16 are dominated by the instrumental fingerprint but also show contaminations by the central emission from the galaxy. Nevertheless, these contaminations did not interfere significantly in the process of instrumental fingerprint removal.

\begin{figure*}
\begin{center}
  \includegraphics[scale=0.65]{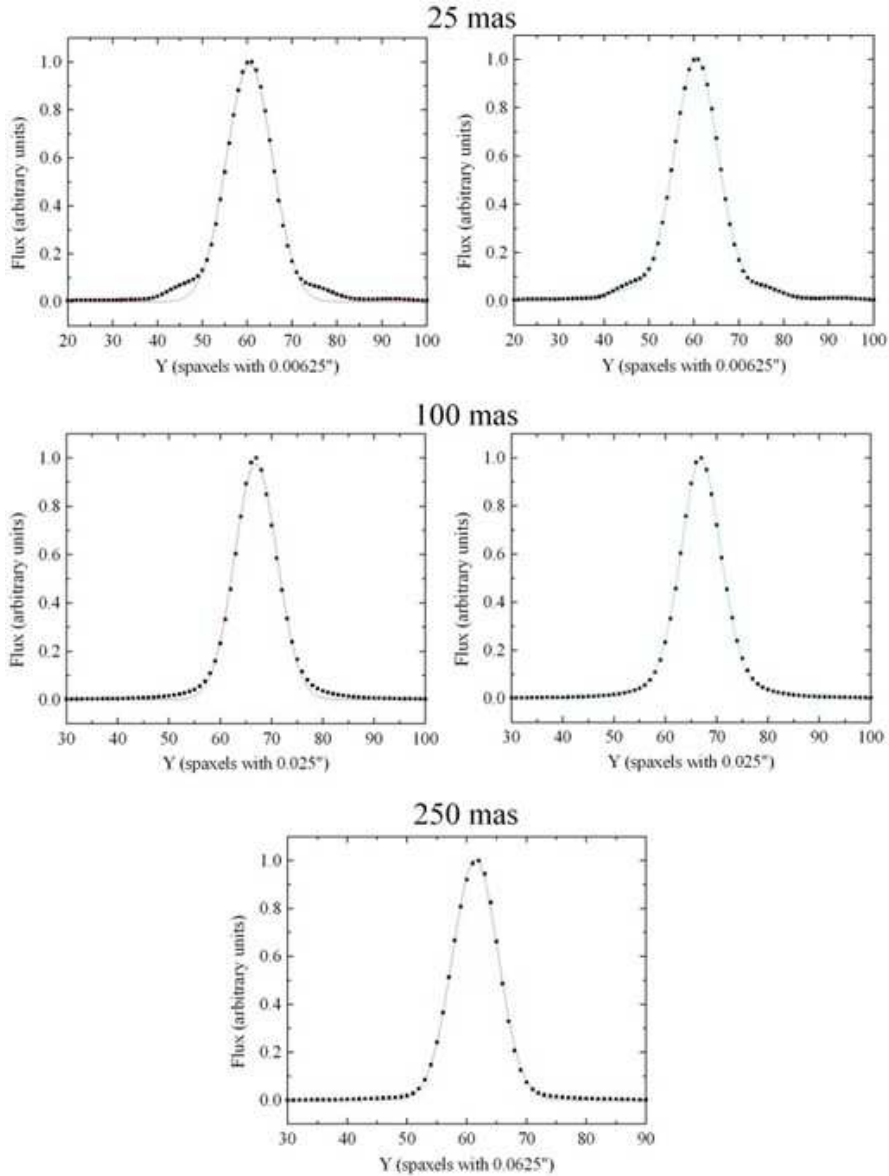}
  \caption{Vertical brightness profiles of images of intermediate-wavelength intervals of the data cubes of (top) HIP 049220, (middle) HIP 048480 and (bottom) HIP 044245. The red curves represent simple Gaussian fits and the green curves represent Gaussian + Lorentzian fits.\label{fig18}}
\end{center}
\end{figure*}

The fractions of the data variances explained by the eigenvectors shown in Fig.~\ref{fig16} indicate that the influence of the fingerprint in the data cubes of Centaurus A decreases from the \textit{J} band to the \textit{K} band. In order to explain that, first of all, it is important to mention that, in our previous experiences \citep{men12}, we have observed that objects with a more intense continuum emission are usually less affected by the instrumental fingerprint than objects with a weaker continuum. Therefore, since the continuum emission from the AGN in Centaurus A increases from the \textit{J} band to the \textit{K} band, the fact that the relevance of the fingerprint decreases from the \textit{J} band to the \textit{K} band is actually expected. On the other hand, since the fingerprint is associated with the spectral continuum and not with the emission lines, objects without continuum emission are not affected by this problem. In data cubes of point-like sources (like HIP 049220, HIP 048480 and HIP 044245), most of the FOV does not show significant continuum emission. As a consequence, in such cases, the instrumental fingerprint normally appears in a subtler way or may not appear at all. This is the reason why we used data cubes of a galaxy to describe the process for instrumental fingerprint removal, which may not be necessary in cases in which the fingerprint is too subtle to be clearly detected.

So far, we were unable to determine the origin of the instrumental fingerprint in SINFONI data cubes. Its characteristics do not seem to change with time. One possibility is that it is caused by some illumination pattern in the instrument \citep{neu07}. However, another possibility is that the fingerprint has been introduced during the data reduction. Nevertheless, our previous experiences \citep{men12} proved that our method is capable of removing this artefact in essentially all cases.

\section{Richardson-Lucy deconvolution}

The last step in our sequence for the treatment of SINFONI data cubes is the Richardson-Lucy deconvolution \citep{ric72, luc74}. A deconvolution is an iterative process that reverts the effects of a convolution. The image of a celestial object observed from the Earth's surface corresponds to the original image convolved with the PSF due to the Earth's atmosphere and to the instrument. Therefore, in the astronomical case, a deconvolution is applied in order to obtain the original image of an object, given the observed image and the PSF of the observation.

\begin{figure*}
\begin{center}
  \includegraphics[scale=0.63]{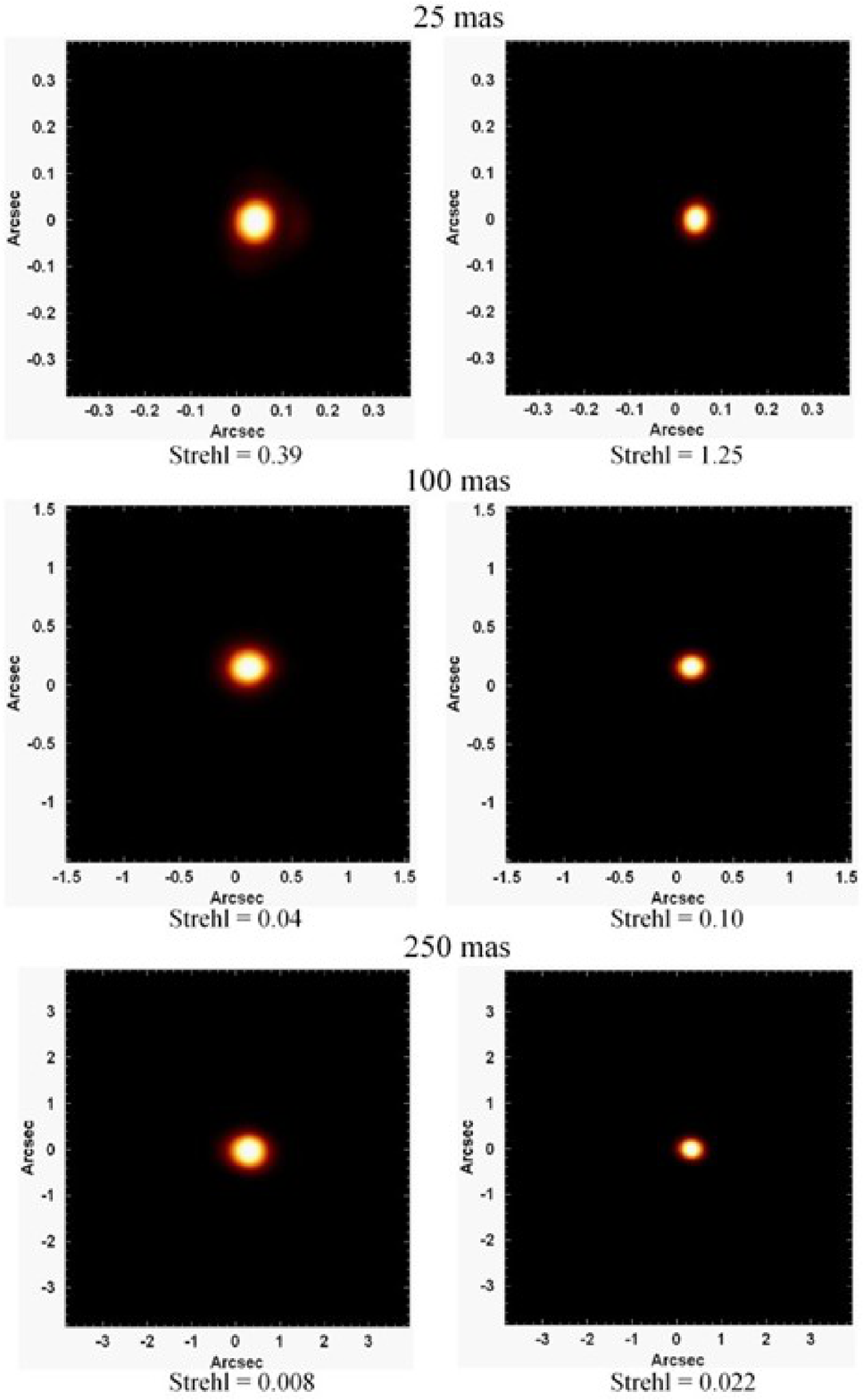}
  \caption{Images of intermediate-wavelength intervals of the data cubes of (top) HIP 049220, (middle) HIP 048480 and (bottom) HIP 044245. The images at left are from the data cubes before the Richardson-Lucy deconvolution and the ones at right are from the data cubes after the Richardson-Lucy deconvolution.\label{fig19}}
\end{center}
\end{figure*}

We verified that, among all the deconvolution techniques that can be applied to astronomical images (Starck \& Murtagh 2006), the Richardson-Lucy is the one that provides the best results for SINFONI data cubes, as it does not introduce detectable artefacts in the images. For more details about the Richardson-Lucy deconvolution, see Paper I. The use of the AO \citep{dav12} usually provides a PSF containing a central diffraction spike, with a shape of an Airy function, and a surrounding halo, with a shape of a Lorentzian function. However, in the case of SINFONI data cubes, the PSFs obtained with the fore-optics with FOVs of 0.8 and 3.2 arcsec can be described by the sum of a Gaussian and a Lorentzian function (equation 13 in Paper I). On the other hand, the PSFs of SINFONI data cubes obtained with the fore-optics with an FOV of 8.0 arcsec (in which the AO is usually not applied) are well described by a simple Gaussian function. 

Similarly to what we discussed in Paper I for NIFS data cubes, it is probable that the existence of a central Gaussian component in the PSFs of SINFONI data cubes (obtained with the two fore-optics with the highest spatial resolutions) is due to the original size of SINFONI spaxels. For an 8 m telescope, the values of the full width at half-maximum (FWHM) of the Airy function at 1.25, 1.65 and 2.2 $\mu$m (the mean wavelengths of the \textit{J}, \textit{H} and \textit{K} bands, respectively) are 0.032, 0.043 and 0.057 arcsec, respectively. For observations taken with the fore-optics with spaxels of 0.125 arcsec $\times$ 0.25 arcsec, all these FWHM values are obviously smaller than one spaxel. So, it would not be possible to resolve the Airy profile, even if the AO was used (which is usually not the case for this fore-optics). One interesting point is that the spatial sampling of this fore-optics is not adequate to precisely resolve even a brightness profile with an FWHM of 0.5 arcsec (which is a typical value of the seeing in this spectral region). Therefore, although the final PSF of an observation taken with this fore-optics has a Gaussian shape, this is probably not a precise reproduction of the brightness profile of the PSF generated by the atmosphere. For observations taken with the fore-optics with spaxels of 0.05 arcsec $\times$ 0.1 arcsec, the FWHM values are smaller than one of the dimensions of the SINFONI spaxels (0.1 arcsec) and very close to the other (0.05 arcsec). As a consequence, one can say that the spatial sampling is still not capable of resolving the Airy profile and this is the reason why a Gaussian profile is usually a good description for the central spike of the PSFs obtained with this fore-optics. For observations taken with the fore-optics with spaxels of 0.0125 arcsec $\times$ 0.025 arcsec, the situation is a little different. The FWHM of the Airy profile in the \textit{J} band is smaller than three spaxels along the horizontal axis and smaller than two spaxels along the vertical axis. Similarly, the FWHM of the Airy profile in the \textit{H} band is smaller than four spaxels along the horizontal axis and smaller than two spaxels along the vertical axis. Finally, the FWHM of the Airy profile in the \textit{K} band is smaller than five spaxels along the horizontal axis and smaller than three spaxels along the vertical axis. So, this fore-optics provides a better spatial sampling for resolving the Airy profile. However, our previous experiences \citep{men12} revealed that, even in this case, the spatial sampling is usually not sufficient to distinguish between a Gaussian profile and an Airy profile and, consequently, the central spike can still be adequately described by a Gaussian function. One should note that the AO corrections work less perfectly in the \textit{J} and \textit{H} bands. As a consequence, in these two bands, it is difficult to reach the diffraction limit and the spaxel scales of 25 and 100 mas are usually adequate to sample the obtained PSFs. Fig.~\ref{fig18} shows the vertical brightness profiles of images of intermediate-wavelength intervals of the data cubes of the standard stars observed with spaxel scales of 25, 100 and 250 mas. Gaussian and Gaussian+Lorentzian fits are also shown.

Fig.~\ref{fig18} confirms that a Gaussian+Lorentzian fit is the most adequate to reproduce the vertical brightness profiles of the stars with spaxel scales of 25 and 100 mas and that a Gaussian fit reproduces, with good precision, the vertical brightness profile of the star with spaxel scale of 250 mas.

Unlike the Airy function, the SINFONI spatial sampling is not wavelength dependent. Therefore, since the PSFs of the observations taken with the fore-optics with FOVs of 0.8 and 3.2 arcsec depend significantly on the spatial sampling (which is not adequate to resolve the Airy profile), these PSFs remain approximately constant along the spectral axis of the data cubes. In the case of observations taken with the fore-optics with an FOV of 8.0 arcsec, as discussed before, although the corresponding PSFs have a Gaussian shape, the spatial sampling is not adequate to precisely reproduce the brightness profile of the PSF generated by the atmosphere. Therefore, the PSFs of observations taken with this fore-optics are also significantly affected by the spatial sampling and, consequently, remain approximately constant along the spectral axis. Considering this stability of the PSFs, we can say that a Richardson-Lucy deconvolution can be applied to SINFONI data cubes (obtained with any fore-optics) using a constant PSF. 

The PSFs of SINFONI data cubes obtained with the fore-optics with FOVs of 0.8 and 3.2 arcsec are considerably difficult to reproduce, as they show a Gaussian and a Lorentzian component with a relative weight that varies from one observation to the other. Due to this complexity and the fact that the Richardson-Lucy deconvolution is very sensitive to the knowledge of the PSF, the best results of the use of this procedure in these two fore-optics are achieved using real PSFs. As explained in Paper I, images of an intermediate-wavelength interval of data cubes of isolated stars or images of the broad wings of permitted lines in type 1 AGNs can be used as PSFs. If a real PSF is not available, one can construct a synthetic PSF as a sum of a Gaussian and a Lorentzian function. In the case of SINFONI data cubes obtained with the fore-optics with FOV of 8.0 arcsec, as explained before, the PSFs have a simple Gaussian shape. So, in these cases, it is relatively easy to construct synthetic PSFs and the Richardson-Lucy deconvolution performed with such PSFs provides results essentially as precise as the ones obtained using real PSFs. For all fore-optics, there are cases in which it is difficult to obtain an estimate of the PSF from the science data cube. In cases like that, it is possible to estimate the PSF from the data cube of the standard star used in the data reduction. This approach, however, should be performed with caution as the PSF of the data cube of the standard star may not be the same as the PSF of the data cube of the observed object, due to differences in the effects of the AO applied to these data cubes and also due to differences in the seeing during these observations. Nevertheless, in many situations, the differences between the PSFs of the science data cube and of the standard star data cube may be subtle and a reliable Richardson-Lucy deconvolution can be applied using the PSF of the standard star data cube. When it is not possible to obtain any estimate of the PSF, the Richardson-Lucy deconvolution should not be applied. As established in Paper I, the Richardson-Lucy gives optimized results if a number of iterations between 6 and 10 is used.

\begin{figure*}
\begin{center}
  \includegraphics[scale=0.60]{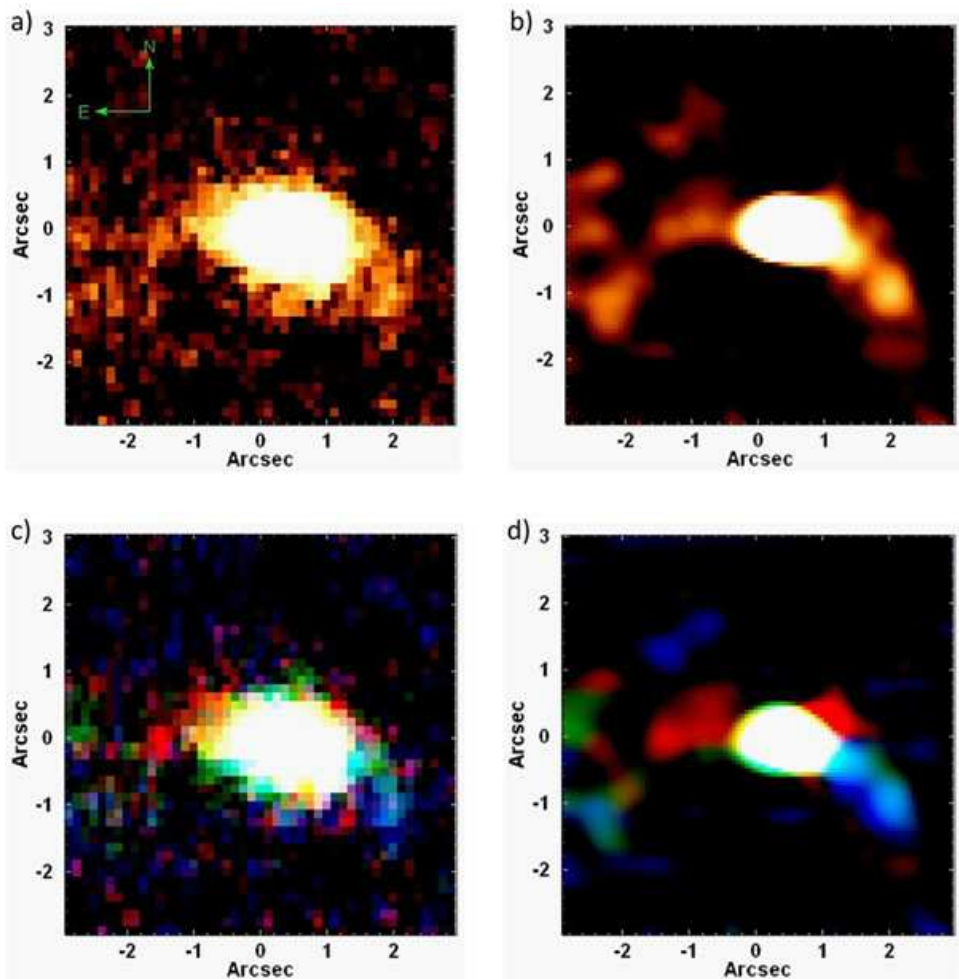}
  \caption{Br$\gamma$ images from the (a) original and (b) treated data cubes of NGC 5643, and the corresponding RGB composite images from the (c) original and (d) treated data cubes of NGC 5643. The colours blue, green and red in panels (c) and (d) correspond to the velocity ranges $-305$ km s$^{-1} \le V_r \le -14$ km s$^{-1}$ , $0$ km s$^{-1} \le V_r \le 28$ km s$^{-1}$ and $42$ km s$^{-1} \le V_r \le 332$ km s$^{-1}$, respectively.\label{fig20}}
\end{center}
\end{figure*}

We applied the Richardson-Lucy deconvolution to the data cubes of the standard stars observed with spaxel scales of 25, 100 and 250 mas, using real PSFs (corresponding to images of the data cubes in intermediate wavelengths) and 10 iterations. Fig.~\ref{fig19} shows images of intermediate-wavelength intervals of the deconvolved and non-deconvolved data cubes of these objects. The Strehl ratios improved from 0.39 to 1.25 in the data cube with spaxel scale of 25 mas, from 0.04 to 0.10 in the data cube with spaxel scale of 100 mas and from 0.008 to 0.022 in the data cube with spaxel scale 250 mas. Without the Richardson-Lucy deconvolution, we usually obtain Strehl ratios in the ranges of 0.3 - 0.7 and 0.01 - 0.2 for data cubes in the fore-optics with FOVs of 0.8 and 3.2 arcsec, respectively. However, with the Richardson-Lucy deconvolution, it is possible to obtain Strehl ratios as high as 0.3 for the fore-optics with FOV of 3.2 arcsec and even higher than 2 for the fore-optics with FOV of 0.8 arcsec \citep{may11}. In the case of the fore-optics with a FOV of 8.0 arcsec, it is difficult to establish a range of values for the Strehl ratio, because these values depend fundamentally on the seeing conditions of the night. The results obtained above indicate that the use of the Richardson-Lucy deconvolution in SINFONI data cubes provides a considerable improvement of the spatial resolution, in any fore-optics. 

\begin{figure*}
\begin{center}
  \includegraphics[scale=0.60]{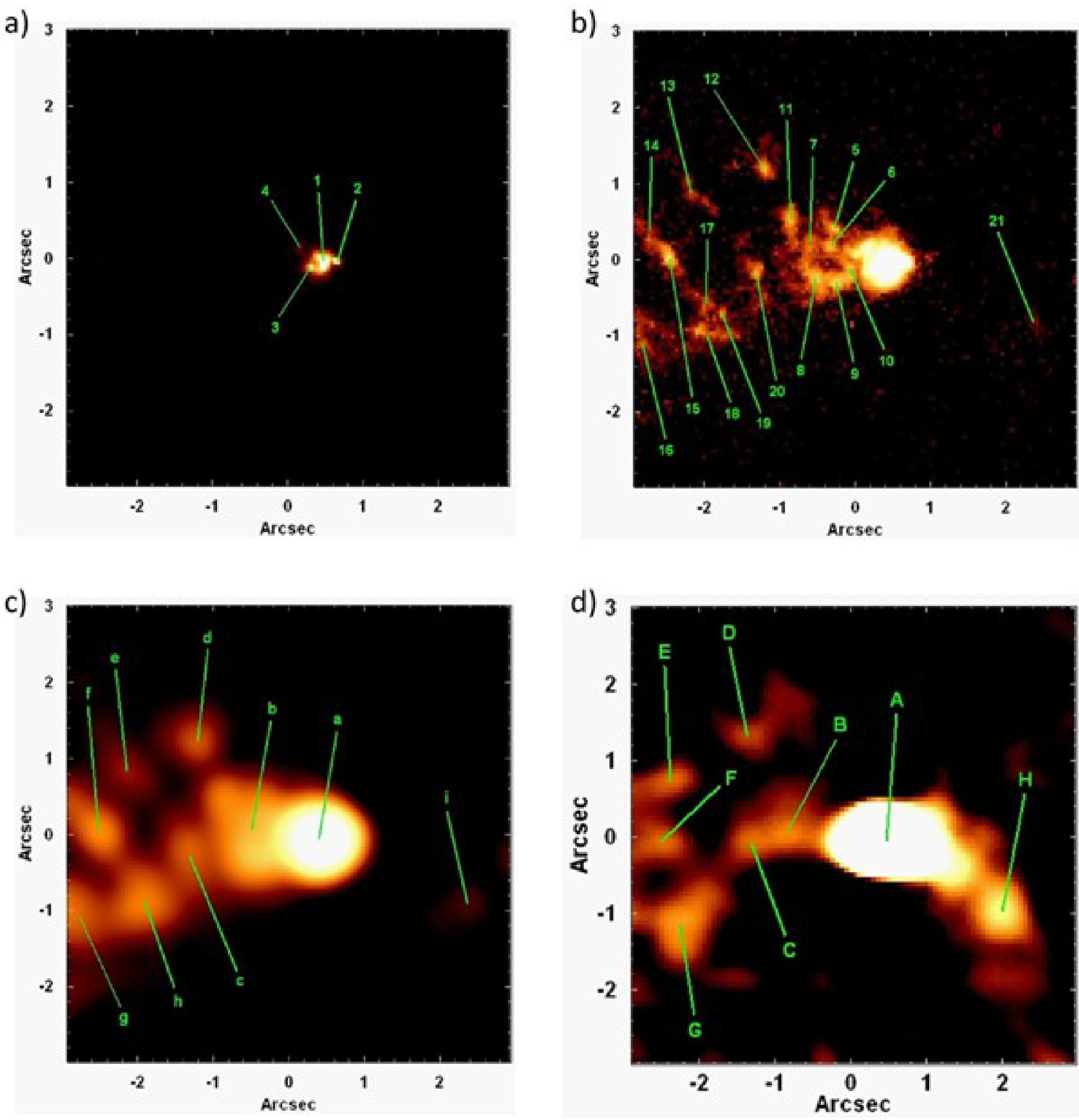}
  \caption{(a) [O III] image of the nuclear region of NGC 5643, obtained with WFPC2 of the \textit{HST}, with the same FOV of the data cube analysed here. (b) The same image shown in (a), with a different look-up table. (c) The same image shown in (a) and (b), after a convolution with an estimate of the PSF of the data cube of NGC 5643. (d) Br$\gamma$ image from the treated data cube of NGC 5643. The main ionized-gas clouds are identified with different notations in (a)/(b), (c) and (d).\label{fig21}}
\end{center}
\end{figure*}

The fact that we obtain Strehl ratios higher than 1 in deconvolved data cubes in the fore-optics with the highest spatial resolution may be somewhat surprising, as, in any observation, this value should always be smaller or equal to 1. Indeed, such high values of the Strehl ratio are only possible here because of the use of deconvolution. Since this procedure is not limited by the diffraction laws as in a real observation, there is, in principle, no upper limit for the Strehl ratio of a deconvolved image.

In our previous experiences \citep{men12}, we have never observed Strehl ratios higher than 0.7 in the fore-optics with an FOV of 0.8 arcsec and higher than 0.2 in the fore-optics with an FOV of 3.2 arcsec. These observed limits are probably related to the size of SINFONI spaxels and not only to limitations in the AO. Following the same strategy used for NIFS data cubes in Paper I, we calculated the Strehl ratios of images containing Airy functions, in the \textit{J}, \textit{H} and \textit{K} bands, sampled in spaxels of 0.0125 arcsec $\times$ 0.025 arcsec and 0.05 arcsec $\times$ 0.1 arcsec (the sizes of the spaxels in the two fore-optics with the highest spatial resolutions). For the fore-optics with spaxels of 0.0125 arcsec $\times$ 0.025 arcsec, the calculated Strehl ratios are $\sim 0.6$, $\sim 0.7$ and $\sim 0.8$, for the \textit{J}, \textit{H} and \textit{K} bands, respectively. On the other hand, for the fore-optics with spaxels of 0.05 arcsec $\times$ 0.1 arcsec, the calculated Strehl ratios are $\sim 0.05$, $\sim 0.09$ and $\sim 0.20$, for the \textit{J}, \textit{H} and \textit{K} bands, respectively. Based on these results, we can say that the observed limits in the Strehl ratios are probably a consequence of the size of SINFONI spaxels.

\section{A scientific example: NGC 5643}

NGC 5643 is an almost face-on ($i \sim 27\degr \pm 5\degr$) SAB(rs)c galaxy. Its nucleus shows a narrow emission-line spectrum (FWHM $\sim$ 300 km s$^{-1}$) of high ionization \citep{san78}. Based on these characteristics, this object was classified as a low luminosity Seyfert 2 by Phillips, Charles \& Baldwin (1983).

Several studies have shown that NGC 5643 seems to be an example of an object containing an obscured active nucleus with anisotropic escape of radiation. This scenario is compatible with the unified model for AGNs. \citet{mor85} analysed velocity maps of optical emission lines, obtained with the imaging Fabri-Perot spectrometer TAURUS, and also flux maps in 6 cm and 20 cm, obtained with the \textit{Very Large Array}. The authors observed an extended optical line emission with an elongated morphology, approximately along the east-west direction (which coincides with the direction of the bar). They also found evidences that the velocity field is similar to the one of models of gas flux in a barred potential. The radio data revealed a nucleus coinciding with the optical nucleus and with the rotation centre, with two lobes (one at each side) along the same direction of the optical emission. Based on these observations, the authors proposed a simple model, in which the gas flows along the bar, forms a disc orthogonal to the major axis of the bar and collimates the radio emission, obscuring, also, the ionizing continuum.

\citet{sch94} analysed images and long-slit spectra, obtained with the \textit{Cerro Tololo Inter-American Observatory}, and verified that the distributions of the emission-line ratios support the hypothesis that the ionizing continuum is collimated in a bicone. \citet{sim97} analysed high resolution images of NGC 5643 obtained with the \textit{Hubble Space Telescope} (\textit{HST}) and observed that the [O III]/H$\alpha$ map has a well defined structure in the form of a `V', which probably corresponds to the projection of a tridimensional cone.

The data cube of NGC 5643 we analyse here is described in further detail in Section 2 and it was treated with all the procedures described before. \citet{hic13}, using standard methodologies, analysed this data cube, together with the data cubes of other nine galaxies (all observed with SINFONI, in the \textit{K} band, with the fore-optics with an FOV of 8.0 arcsec), forming a sample containing five Seyfert galaxies and five quiescent galaxies. The authors verified that the Seyferts in the sample have a more centrally concentrated nuclear stellar surface brightness, a lower stellar velocity dispersion within a radius of 200 pc, an elevated H$_2$ 1-0 S(1) luminosity out to a radius of at least 250 pc and a more centrally concentrated H$_2$ surface brightness. The authors interpreted these characteristics of the Seyfert galaxies as an indication for the existence of a thermally cold (in comparison to the bulge) nuclear structure, composed of a gas reservoir and a relatively young stellar population, in these objects. This indicates that the fuelling of a Seyfert galaxy is associated with the formation of the dynamically cold nuclear component. In this work, we are reanalysing the data cube of NGC 5643 with our data treatment methodology, showing different scientific aspects. It is important to emphasize that none of the results obtained with our analysis of the data cube of NGC 5643 is in conflict with what was obtained by \citet{hic13}, as the work made by these authors was focused on a comparison between Seyfert and inactive galaxies, not on the details of individual galaxies. Our analysis here is only about NGC 5643 and focused on the emission and kinematics of the ionized and molecular gas.

We compared images from the treated and non-treated data cubes of NGC 5643. The non-treated data cube corresponds to the median of the five data cubes obtained after the data reduction. Only the correction of the DAR was applied to each one of these five data cubes. Fig.~\ref{fig20} shows images of the Br$\gamma$ emission line (with the adjacent spectral continuum subtracted) from the treated and non-treated data cubes of NGC 5643. The corresponding RGB composite images (based on the radial velocity intervals) are also shown. All the images show the existence of an extended emission approximately along the east-west direction. However, only the images from the treated data cube allow a clear visualization of individual ionized-gas clouds, which are almost undetectable in the images from the non-treated data cube. The extended emission along the east-west direction represents the narrow-line region (NLR), in the form of a bicone, of this galaxy. This result is compatible with the observations made by \citet{mor85} and Schmitt et al. (1994). 

\begin{table*}
\begin{center}
\caption{Comparison between the central ionized-gas clouds of NGC 5643 identified using SINFONI data and \textit{HST} data.\label{tbl2}}
\begin{tabular}{cccc}
\hline
Identification of the cloud      & Identification of the cloud          &  Identification of the cloud       & Radial velocity obtained               \\
in the Br$\gamma$ image       & in the [O III] image convolved   &  in the original [O III] image     & using the Br$\gamma$ emission   \\
                                               & with an estimate of the PSF       &                                                  & line (km s$^{-1}$)                        \\
                                               & of the data cube of NGC 5643   &                                                  &                                                      \\
\hline
\textit{B}  & \textit{b} & 5+6+7+8+9+10+11 & $187 \pm 18$ \\
\textit{C}  & \textit{c} & 20 & $184 \pm 5$ \\
\textit{D}  & \textit{d} & 12 & $-132 \pm 6$ \\
\textit{E}  & \textit{e} ? & 13 ? & $44 \pm 9$ \\
\textit{F}  & \textit{f} & 14+15 & $43 \pm 2$ \\
\textit{G}  & \textit{g+h} ?& 16+17+18+19 ? & $-43 \pm 26$ \\
\textit{H}  & \textit{i} ? & 21 ? & $-111 \pm 32$ \\
\hline
\end{tabular}
\end{center}
\end{table*}

\begin{figure}
\begin{center}
  \includegraphics[scale=0.30]{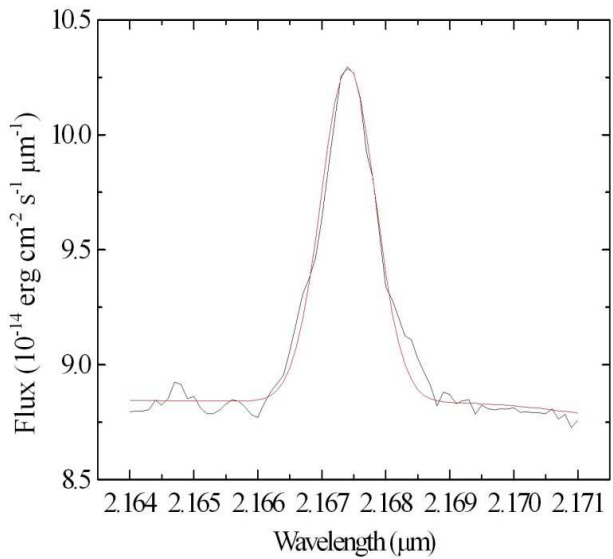}
  \caption{Br$\gamma$ emission line of the spectrum of the ionized-gas cloud \textit{C} identified in the treated data cube of NGC 5643. A Gaussian fit is shown in red.\label{fig22}}
\end{center}
\end{figure}

Fig.~\ref{fig21} shows an [O III] image with two different look-up tables of the nuclear region of NGC 5643, obtained with the Wide-Field Planetary Camera 2 (WFPC2) of the \textit{HST} (this image was previously analysed by Simpson et al. 1997). Since the spatial resolution of the \textit{HST} is considerably higher than the spatial resolution of the data cube of NGC 5643, it is difficult to establish a comparison between the [O III] image and the Br$\gamma$ image from the treated data cube. Therefore, in order to make that comparison, we convolved the [O III] image with an estimate of the PSF of the data cube. The result of this convolution and the Br$\gamma$ image from the treated data cube of NGC 5643 are also shown in Fig.~\ref{fig21}. The main ionized-gas clouds are identified in all the images in Fig.~\ref{fig21} (and are compared in Table~\ref{tbl2}).

The first thing we can notice in Fig.~\ref{fig21} is that the [O III] image reveals the existence of an one-sided ionization cone \citep{sim97}, as the ionized-gas clouds extend mainly to the east, but not to the west of the nucleus. Only one very weak ionized-gas cloud (21/i) is located at the east of the nucleus. On the other hand, the Br$\gamma$ image shows a bicone. The difference between these two images is probably due to the extinction by dust, which affects more the [O III] image than the Br$\gamma$ image. Indeed, Schmitt et al. (1994) concluded that there is an increase of the interstellar extinction to the west of the nucleus and a decrease of the interstellar extinction to the east. This behaviour is certainly compatible with the existence of an apparent one-sided ionization cone at the east of the nucleus, in the [O III] image. 

By analysing the positions of the ionized-gas clouds in Fig.~\ref{fig21}, we can see a good correspondence between the [O III] and the Br$\gamma$ images. There are, however, a few discrepancies that should be discussed. Clouds \textit {g} and \textit{h} in the [O III] image, for example, appear as cloud \textit{G} only in the Br$\gamma$ image. Clouds \textit{e} and \textit{E} appear to be associated with each other but their positions are a little different. The same happens to clouds \textit{i} and \textit{H}. These discrepancies may be related to the extinction by dust. However, it is important to mention that the [O III] emitting areas do not necessarily correspond to the Br$\gamma$ emitting areas. Therefore, the discrepancies between the [O III] and the Br$\gamma$ images in Fig.~\ref{fig21} may represent real differences between the [O III] and the Br$\gamma$ emitting areas. Table~\ref{tbl2} shows a comparison between the identifications of the ionized-gas clouds in all the images in Fig.~\ref{fig21}.

We extracted spectra from circular regions, with a radius of 0.2 arcsec, centred on all the clouds of the NLR visible in the Br$\gamma$ image from the treated data cube. We then shifted all the spectra to the rest frame, using the task `\textit{dopcor}' (from the {\sc `noao'} package) with \textit{z} = 0.003999 (NASA Extragalactic Database - NED), and applied a correction due to the heliocentric velocity, using the task `\textit{rvcorrect}' (from the {\sc `astutil'} package). After that, we fitted Gaussian functions to the Br$\gamma$ emission lines in the obtained spectra, in order to determine the radial velocity values of the ionized-gas clouds. This entire analysis was performed in {\sc iraf} environment. Fig.~\ref{fig22} shows the Gaussian fit applied to the Br$\gamma$ emission line of the spectrum of the ionized-gas cloud \textit{C}. The profile of the emission line is well described by the Gaussian function and, therefore, we conclude that the fit provides a reliable value for the corresponding radial velocity. The radial velocity values of the ionized-gas clouds are shown in Table~\ref{tbl2}. A velocity map of the Br$\gamma$ emission line would give more details about the kinematics of the ionized gas in the nuclear region of NGC 5643. However, we verified that the S/N ratio of the Br$\gamma$ emission line in the regions where most of the ionized-gas clouds are located is considerably low. As a consequence, the obtained velocity values are not sufficiently precise to construct a velocity map. This is the reason why we determined the velocity values of the Br$\gamma$ line for each ionized-gas cloud using spectra extracted from circular regions, centred on each cloud.

\begin{figure*}
\begin{center}
  \includegraphics[scale=0.60]{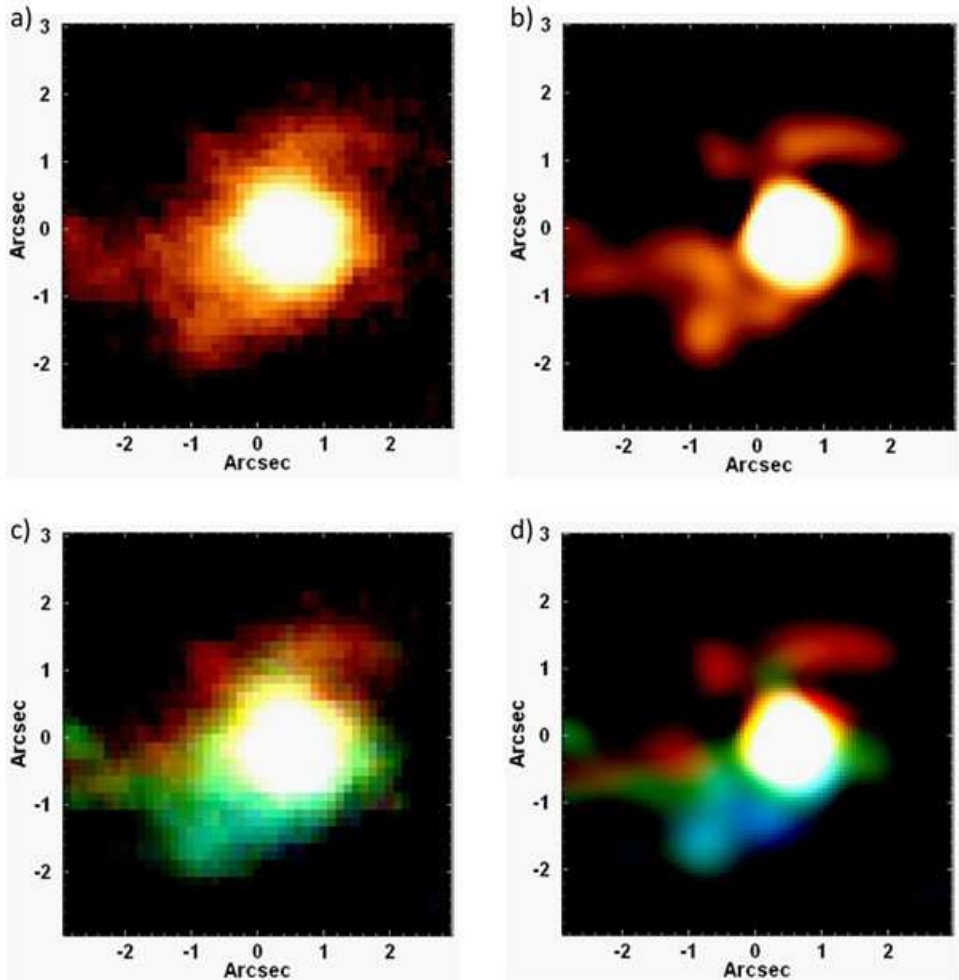}
  \caption{H$_2 \lambda 21218$ images from the (a) original and (b) treated data cubes of NGC 5643, and the corresponding RGB composite images from the (c) original and (d) treated data cubes of NGC 5643. The colours blue, green and red in panels (c) and (d) correspond to the velocity ranges $-345$ km s$^{-1} \le V_r \le -48$ km s$^{-1}$ , $-34$ km s$^{-1} \le V_r \le 0$ km s$^{-1}$ and $8$ km s$^{-1} \le V_r \le 305$ km s$^{-1}$, respectively.\label{fig23}}
\end{center}
\end{figure*}

Table~\ref{tbl2} shows that the moduli of the radial velocities of the three ionized-gas clouds with the highest projected distances from the central AGN (clouds \textit{E}, \textit{F} and \textit{G}) are considerably lower than the moduli of the radial velocities of the other clouds. We cannot perform a detailed analysis of the ionized-gas kinematics due to our limited FOV. However, the contrast between the radial velocities of clouds \textit{E}, \textit{F} and \textit{G} and the others suggests the existence of an acceleration of the ionized gas near the AGN, followed by a deceleration due to the interaction with the interstellar medium. A similar behaviour was also observed in the NIFS data cube of the nuclear region of the Seyfert 1 galaxy NGC 4151 (see Paper I).

In order to analyse the molecular emission of H$_2$ at the surroundings of the AGN in NGC 5643, we constructed images of the H$_2 \lambda 21218$ emission line (with the adjacent spectral continuum subtracted) from the treated and non-treated data cubes of this galaxy. Fig.~\ref{fig23} shows the obtained images and also RGB composite images based on the radial velocity intervals. Similarly to what was observed in the images of the Br$\gamma$ emission line in Fig.~\ref{fig20}, all the H$_2 \lambda 21218$ images in Fig.~\ref{fig23} show a clear extended emission; however, only in the images from the treated data cube we can identify individual molecular-gas clouds. Besides that, we can see that the highest velocity values of the molecular gas can be found approximately along the north-south direction, which is perpendicular to the direction of the ionization cone (east-west). The images in Fig.~\ref{fig23} may give the impression that part of the extended emission has been taken away by the data treatment; however, this is just a consequence of the improvement of the spatial resolution due to the use of the Richardson-Lucy deconvolution.

\begin{figure}
\begin{center}
  \includegraphics[scale=0.68]{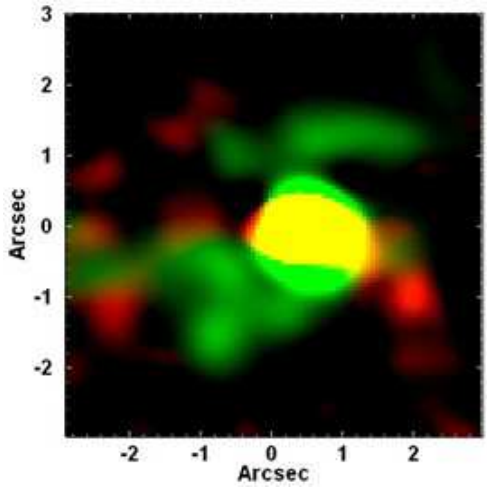}
  \caption{RG composite image from the treated data cube of NGC 5643, with the Br$\gamma$ emission in red and the H$_2 \lambda 21218$ emission in green.\label{fig24}}
\end{center}
\end{figure}

\begin{figure}
\begin{center}
  \includegraphics[scale=0.69]{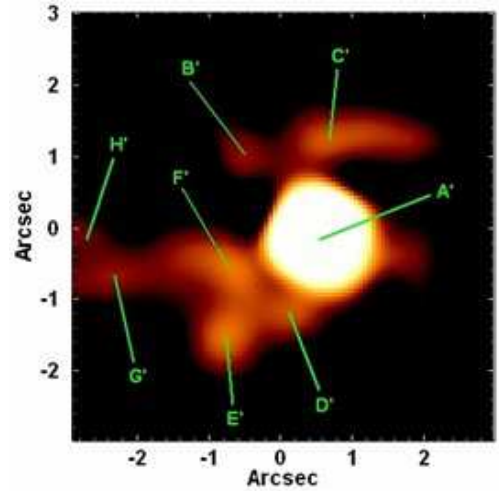}
  \caption{H$_2 \lambda 21218$ image from the treated data cube of NGC 5643, with the main molecular-gas clouds identified.\label{fig25}}
\end{center}
\end{figure}

\begin{figure}
\begin{center}
  \includegraphics[scale=0.22]{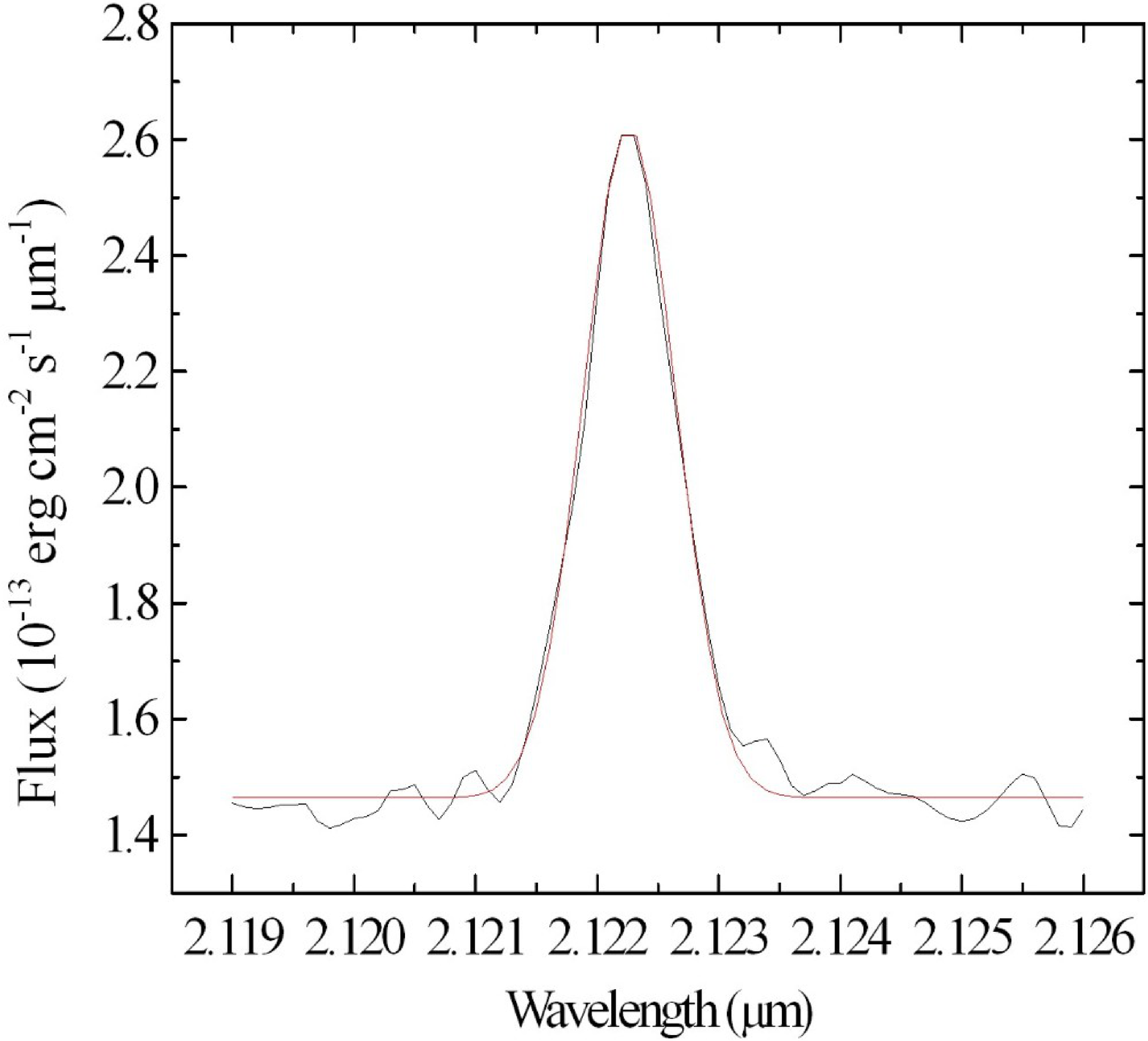}
  \caption{H$_2 \lambda 21218$ emission line of the spectrum of the molecular-gas cloud \textit{C'} identified in the treated data cube of NGC 5643. A Gaussian fit is shown in red.\label{fig26}}
\end{center}
\end{figure}

A model capable of explaining the observed morphology of the molecular gas in Fig.~\ref{fig23} involves the existence of a molecular torus/disc structure along the north-south direction. One possibility is that the gas flows along the torus/disc towards the nucleus, feeding the AGN. This torus/disc collimates the AGN emission along the ionization cone. This scenario is compatible with the models proposed by \citet{mor85}, Schmitt et al. (1994) and \citet{sim97} and also with the unified model for AGNs. Fig.~\ref{fig24} shows an RG composite image with the Br$\gamma$ emission in red and the H$_2 \lambda 21218$ emission in green. This image makes it easier to visualize the proposed scenario involving a molecular torus/disc feeding the AGN and collimating the AGN emission. One important point that should be discussed here is the fact that Fig.~\ref{fig24} shows the existence of some molecular gas in areas corresponding to the ionization cone of NGC 5643. Considering that most of the H$_2$ molecules should be destroyed along the ionization cone, it is probable that the superposition of emission from ionized and molecular gas we are seeing in Fig.~\ref{fig24} is caused by the projection on the plane of the sky and the corresponding emitting areas are actually not coplanar. A simple test for that hypothesis would be to compare the velocity values of the ionized-gas clouds and molecular-gas clouds that are apparently superposed. If the values are compatible, then a reasonable assumption is that the superposed areas are associated with each other. 
\begin{table*}
\begin{center}
\caption{Radial velocities calculated for the main molecular-gas clouds identified in the H$_2 \lambda 21218$ image from the treated data cube of NGC 5643.\label{tbl3}}
\begin{tabular}{cc}
\hline
Identification of the cloud      & Radial velocity obtained               \\
in the H$_2 \lambda 21218$  & using the H$_2 \lambda 21218$  \\
image                                      & emission line (km s$^{-1}$)         \\
\hline
\textit{B'}  & $89 \pm 2$ \\
\textit{C'}  & $42 \pm 2$ \\
\textit{D'}  & $-57 \pm 2$ \\
\textit{E'}  & $-47 \pm 2$ \\
\textit{F'}  & $-17 \pm 5$ \\
\textit{G'}  & $28 \pm 3$ \\
\textit{H'}  & $-11 \pm 4$ \\
\hline
\end{tabular}
\end{center}
\end{table*}
On the other hand, if the values are not compatible, then it is very unlikely that the superposed areas are related to each other and the apparent superpositions are probably just caused by the projection on the plane of the sky. In order to perform this test, we extracted the spectra from circular regions, with a radius of 0.2 arcsec, centred on each molecular-gas cloud detected in the H$_2 \lambda 21218$ image from the treated data cube of NGC 5643. The identifications of the molecular-gas clouds in the H$_2 \lambda 21218$ image are shown in Fig.~\ref{fig25}. After that, we repeated the procedure used with the spectra extracted from the ionized-gas clouds: we shifted all the spectra to the rest frame (using the task `\textit{dopcor}' from the {\sc `noao'} package, with \textit{z} = 0.003999) and applied a correction due to the heliocentric velocity (using the task `\textit{rvcorrect}' from the {\sc `astutil'} package). Then, we fitted Gaussian functions to the H$_2 \lambda 21218$ emission lines, in order to determine the radial velocities of the molecular-gas clouds. Fig.~\ref{fig26} shows the Gaussian fit applied to the H$_2 \lambda 21218$ of the spectrum of the molecular-gas cloud \textit{C'}. The radial velocity value provided by this fit is essentially as precise as the one provided by the Gaussian fit applied to the Br$\gamma$ emission line in Fig.~\ref{fig22}. The radial velocities obtained for the molecular-gas clouds are shown in Table~\ref{tbl3}.

Table~\ref{tbl3} confirms that the molecular-gas clouds with the highest velocity values are located approximately along the north-south direction. Figs. 21, 24 and 25 reveal that there is a partial superposition between clouds \textit{C} and \textit{F'} and between clouds \textit{F} and \textit{H'}. No more significant superpositions can be observed. However, Tables 2 and 3 show that the velocity values of clouds \textit{C} and \textit{F} are not compatible (at 3$\sigma$ level) with the velocity values of clouds \textit{F'} and \textit{H'}, respectively. This indicates that, as expected, the molecular emission of H$_2$ is not coplanar with the ionized-gas emission along the ionization cone of the NGC 5643. 

\section{Summary and conclusions}

We presented a treatment procedure for VLT/SINFONI data cubes, which includes the following steps: correction of the DAR; spatial re-sampling, which provides a better visualization of the contours of the spatial structures; Butterworth spatial filtering, to remove high spatial-frequency components from the images of the data cubes; instrumental fingerprint removal; Richardson-Lucy deconvolution, to improve the spatial resolution of the data cubes.

The main conclusions related to the steps of the data treatment procedure are as follows. 

\begin{itemize}

\item The DAR effect may be significant for SINFONI data cubes obtained with the fore-optics with FOVs of 0.8 and 3.2 arcsec, even if only one spectral band is considered. In the case of the fore-optics with FOV of 8.0 arcsec, the correction of the DAR may not be necessary. Both the theoretical and the practical approaches are very effective in removing the DAR, although the practical approach is a little more precise.

\item The spatial re-sampling followed by an interpolation of the values provides a clearer visualization of the contours of the spatial structures. We usually apply the spatial re-sampling to SINFONI data cubes (observed with any fore-optics) in order to double the number of spaxels along the horizontal and vertical axis.

\item A Butterworth spatial filtering using a filter corresponding to the product between two identical circular filters removes most of the high spatial-frequency components (mainly in the form of horizontal stripes) from SINFONI data cubes, without compromising the images.

\item An instrumental fingerprint in the form of a large horizontal stripe at the bottom of the images and with a characteristic spectral signature is usually found in SINFONI data cubes. Our method involving the PCA Tomography technique is very effective in removing this artefact from the data cubes.

\item With the Richardson-Lucy deconvolution, it is possible to obtain Strehl ratios as high as 0.3 for the fore-optics with an FOV of 3.2 arcsec and even higher than 2 for the fore-optics with an FOV of 0.8 arcsec. The PSFs of SINFONI data cubes correspond to a simple Gaussian function for observations obtained with the fore-optics with an FOV of 8.0 arcsec. On the other hand, the PSFs are composed of a Gaussian and a Lorentzian component for observations obtained with the fore-optics with FOVs of 0.8 and 3.2 arcsec.

\end{itemize}

In order to give a scientific example of the efficacy of our data treatment procedure, we analysed a data cube of the nuclear region of the galaxy NGC 5643, obtained with the fore-optics with an FOV of 8.0 arcsec, and compared the results obtained with and without the treatment procedure. The main conclusions of this analysis are the following.

\begin{itemize}

\item An extended emission can be easily seen in the images of the Br$\gamma$ emission line from the treated and non-treated data cubes of NGC 5643. However, only the image from the treated data cube allows a clear visualization of individual ionized-gas clouds, which are almost impossible to visualize in the image from the non-treated data cube.

\item Although we cannot perform a detailed analysis of the ionized-gas kinematics (due to our limited FOV), the radial velocities determined for the ionized-gas clouds suggest the existence of an acceleration of the ionized gas near the AGN, followed by a deceleration due to the interaction with the interstellar medium.

\item An extended molecular emission can also be seen in the images of the H$_2 \lambda 21218$ emission line from the treated and non-treated data cubes of NGC 5643. However, only the image from the treated data cube, again, allows a clear visualization of individual molecular-gas clouds.

\item The morphology of the H$_2$ molecular emission can be explained by the existence of a molecular torus/disc structure along the north-south direction. One hypothesis is that the gas flows along the torus/disc towards the nucleus, feeding the AGN. This torus/disc collimates the AGN emission along the ionization cone. This scenario is compatible with the models proposed by \citet{mor85}, Schmitt et al. (1994) and \citet{sim97} and also with the unified model for AGNs.

\end{itemize}

Considering the benefits provided by our data treatment procedure, we conclude that this methodology may improve the quality of the analysis of SINFONi data cubes. The scripts of all procedures included in our method can be found at \textit{http://www.astro.iag.usp.br/$\sim$PCAtomography}.

\section*{Acknowledgements}

This work is based on observations obtained with VLT/SINFONI (Eisenhauer et al. 2003; Bonnet et al. 2004). We would like to thank FAPESP for support under grants 2012/02268-8 (RBM) and 2012/21350-7 (TVR) and also an anonymous referee for valuable comments about this paper.

\end{document}